\documentclass[pra,twocolumn,showpacs,superscriptaddress,aps]{revtex4}
\usepackage{amsfonts}
\usepackage{amsmath}
\usepackage{amssymb}
\usepackage{graphicx,color}
\usepackage{dcolumn}
\usepackage{times}

\DeclareMathOperator{\sech}{sech}

\begin{document}

\title{Dark solitons near potential and nonlinearity steps}

\author{F. Tsitoura}
\affiliation{Department of Physics, University of Athens,
Panepistimiopolis, Zografos, Athens 15784, Greece}

\author{Z. A. Anastassi}
\affiliation{Department of Mathematics, Statistics and Physics,
College of Arts and Sciences, Qatar University, 2713 Doha, Qatar}

\author{J. L. Marzuola}
\affiliation{Department of Mathematics, University of North Carolina,
Chapel Hill, NC 27599, USA}

\author{P. G. Kevrekidis}
\affiliation{Department of Mathematics and Statistics, University of Massachusetts,
Amherst, Massachusetts 01003-4515 USA}

\affiliation{Center for Nonlinear Studies and Theoretical Division, Los Alamos
National Laboratory, Los Alamos, NM 87544}

\author{D. J. Frantzeskakis}
\affiliation{Department of Physics, University of Athens,
Panepistimiopolis, Zografos, Athens 15784, Greece}

\begin{abstract}

We study dark solitons near potential and nonlinearity steps 
and combinations thereof, forming rectangular barriers. 
This setting is relevant to the contexts 
of atomic Bose-Einstein condensates (where such steps
can be realized by using proper external fields) and nonlinear optics 
(for beam propagation near interfaces separating optical media of 
different refractive indices). We use perturbation theory to develop 
an equivalent particle theory, describing the matter-wave or optical 
soliton dynamics as the motion of a particle in
an effective potential. This Newtonian dynamical problem provides information for 
the soliton statics and dynamics, including scenarios of reflection, transmission, 
or quasi-trapping at such steps. 
The case of multiple such steps and its connection to barrier potentials
is also touched upon. 
Our analytical predictions are found to be
in very good agreement with the corresponding numerical results.

\end{abstract}


\maketitle

\section{Introduction}

The interaction of solitons with impurities is a fundamental problem that
has been considered in various branches of physics -- predominantly in
nonlinear wave theory \cite{RMP} and solid state physics \cite{kos1} -- as well as
in applied mathematics (see, e.g., recent work \cite{delta_cauchy} and references therein).
Especially, in the framework of the nonlinear Schr\"{o}dinger (NLS) equation,
the interaction of bright and dark solitons with $\delta$-like
impurities has been investigated in many works
(see, e.g., Refs.~\cite{kos2,cao,holmes,holmer,vvk1}). Relevant studies
in the physics of atomic Bose-Einstein condensates (BECs) have also been performed
(see, e.g., Refs.~\cite{gt,nn,nn2,greg,sch}), as well as in settings involving
potential wells~\cite{pantofl,pantofl1} and barriers~\cite{gardiner,martin}
(see also Ref.~\cite{nogami} for earlier work in a similar model).
In this context, localized impurities can be created as focused far-detuned
laser beams, and have already been used in experiments involving dark solitons
\cite{engels,hulet}. Furthermore, experimental results on the scattering of
matter-wave bright solitons on Gaussian barriers in either $^7$Li \cite{randy}
or $^{85}$Rb \cite{dic} BECs have been reported as well. More recently, such
soliton-defect interactions were also explored in the case of multi-component BECs and dark-bright solitons,
both in theory \cite{vaspra} and in an experiment \cite{exp_eng}.

On the other hand, much attention has been paid to
BECs with spatially modulated
interatomic interactions, so-called ``collisionally inhomogeneous condensates''
\cite{g,fr_inhom}; for a review with a particular bend
towards periodic such interactions see also Ref.~\cite{borya}). Relevant studies
in this context have explored a variety of interesting phenomena:
these include, but are not limited to adiabatic compression of matter-waves 
\cite{g,AS}, Bloch oscillations of solitons \cite{g}, 
emission of atomic solitons \cite{in4,fot_scat},
scattering of matter waves through barriers \cite{in5},
emergence of instabilities of solitary waves
due to periodic variations in the scattering length~\cite{kody},
formation of stable condensates exhibiting both attractive and repulsive
interatomic interactions \cite{in8}, solitons in combined linear and nonlinear
potentials \cite{lnl1,lnl2,kominis,MJS1,MJS2}, 
generation of solitons \cite{21} and vortex rings \cite{22},
control of Faraday waves \cite{alex}, vortex dipole dynamics in spinor BECs \cite{spin},
and others.

Here, we consider a combination of the above settings, namely we
consider a one-dimensional (1D) setting involving potential and nonlinearity steps,
as well as pertinent rectangular barriers,
and study statics, dynamics and scattering of dark solitons.
In the BEC context, recent experiments have demonstrated robust dark solitons in the quasi-1D
setting \cite{dsexp}. In addition, potential steps in BECs can be realized 
by trapping potentials featuring piece-wise constant profiles 
(see, e.g., Refs.~\cite{step_carr,carr_rect} and discussion in the
next Section). Furthermore, nonlinearity steps can be realized too, upon employing
magnetically \cite{FRM} or optically \cite{FRM2} induced Feshbach resonances,
that can be used to properly tune the interatomic interactions strength
-- see, e.g., more details in Refs.~\cite{lnl2,fot_scat} and discussion in the next Section.

Such a setting involving potential and nonlinearity steps, finds also applications
in the context of nonlinear optics.
There, effectively infinitely long potential and nonlinearity steps
of constant and finite height, describe interfaces separating optical media characterized by
different linear and nonlinear refractive indices \cite{molnew}. In such settings, it has been shown
\cite{newell_two,kikos,kivqu,kominis2} that the dynamics of self-focused light channels --
in the form of spatial bright solitons -- can be effectively described
by the motion of an equivalent particle in
effective step-like potentials. 
This ``equivalent particle theory'' actually corresponds to the adiabatic
approximation of the perturbation theory of solitons \cite{RMP}, while reflection-induced
radiation effects can be described at a higher-order approximation \cite{kikos,kivqu}.
Note that similar studies, but for dark solitons in settings involving
potential steps and rectangular barriers, have also been performed -- see, e.g.,
Ref.~\cite{saktam} for an effective particle theory, and Refs.~\cite{prouk,analogies,sak_pot}
for numerical studies of reflection-induced radiation effects. However,
to the best of our knowledge, the statics and dynamics of dark solitons
near potential and nonlinearity steps, have not been systematically
considered so far in the literature, although a special
version of such a setting
has been touch upon in Ref.~\cite{lnl2}.

It is our purpose, in this work, to address this problem. In particular,
our investigation and a description of our presentation is as follows. First, in Sec.~II,
we provide the description and modeling of the problem; although this is done
in the context of atomic BECs, our model can straightforwardly be used for similar
considerations in the context of optics, as mentioned above. In the same Section, we
apply perturbation theory for dark solitons to show that, in the adiabatic approximation,
soliton dynamics is described by the motion of an equivalent particle in an effective
potential. The latter has a tanh-profile, but -- in the presence of the nonlinearity step --
can also exhibit an elliptic and a hyperbolic fixed point. We show that stationary soliton
states do exist at the fixed points of the effective potential, but are unstable 
(albeit in different ways, as is explained below) according to
a Bogoliubov-de Gennes (BdG) analysis \cite{pita,book_fr} that we perform;
we also use an analytical approximation to derive the unstable eigenvalues
as functions of the magnitudes of the potential/nonlinearity steps.
In Sec.~III we study the soliton dynamics for various parameter
values, pertaining to different forms of the effective potential, including the case of
rectangular barriers formed by combination of adjacent potential and nonlinearity steps.
Our numerical results -- in both statics and dynamics -- are found to be in very good
agreement with the analytical predictions.
We also investigate the possibility of soliton trapping in the
vicinity of the hyperbolic fixed point of the
effective potential; note that such states could be characterized as 
``surface dark solitons'', as they are formed at linear/nonlinear interfaces separating
different optical or atomic
media. We show that quasi-trapping of solitons is possible,
in the case where nonlinearity steps are present; the pertinent (finite) trapping time is found
to be of the order of several hundreds of milliseconds, which suggests that such soliton quasi-trapping
could be observable in real BEC experiments. Finally, in Sec.~IV 
we summarize our findings,
discuss our conclusions, and provide provide perspectives for future studies.

\section{Model and analytical considerations}

\subsection{Setup}

As noted in the Introduction, our formulation originates from the context of atomic BECs
in the  mean-field picture \cite{pita}. We thus consider a quasi-1D setting whereby matter
waves, described by the macroscopic wave function $\Psi (x,t)$, are oriented along the
$x$-direction and are confined in a strongly anisotropic (quasi-1D) trap. The latter,
has the form of a rectangular box of lengths $L_{x}\gg L_{y}=L_{z}\equiv L_{\perp }$,
with the transverse length $L_{\perp}$ being on the order of the healing length $\xi$.
Such a box-like trapping potential, $V_b(x)$, can be approximated by a super-Gaussian function,
of the form:
\begin{equation}
V_{b}(x)=V_{0}\left[1-\mathrm{\exp }\left( -\left( \frac{x}{w}%
\right)^{\gamma}\right) \right],
\end{equation}
where $V_{0}$ and $w$ denote the trap amplitude and width, respectively.
The particular value of the exponent $\gamma \gg 1$ is not especially
important; here we use $\gamma=50$.
In this setting, our aim is to consider dark solitons near potential and nonlinearity steps,
located at $x=L$. To model such a situation, we start from the Gross-Pitaevskii (GP)
equation \cite{pita,book_fr}:
\begin{eqnarray}
i\hbar\frac{\partial \Psi}{\partial t} &=& \Big[-\frac{\hbar^2}{2m} \partial_x^2
+  g(x) |\Psi|^2 + V(x) \Big] \Psi,
\label{gpe1}
\end{eqnarray}
Here, $\Psi(x,t)$ is the mean-field wave function, $m$ is the atomic mass,
$V(x)$ represents the external potential, while $g_{1D}(x)=(9/4L_{\perp}^2)g_{3D}$
is the effectively 1D interaction strength, with $g_{3D}=4\pi \hbar^2 \alpha(x)/m$ being its 3D counterpart
and $\alpha(x)$ being the scattering length
(assumed to be $\alpha>0$, $\forall x$, corresponding to repulsive interatomic interactions).
The external potential and the scattering length are then taken to be of the form:
\begin{eqnarray}
V(x)&=&V_b(x)+
\begin{cases}
V_{\rm{L}},& x<L\\
V_{\rm{R}},&  x>L
\end{cases},~ \\
\alpha(x)&=&
\begin{cases}
\alpha_{\rm{L}},& x<L \\
\alpha_{\rm{R}},&  x>L
\end{cases},
\label{setting}
\end{eqnarray}
%
where $V_{{\rm L,R}}$ and $\alpha_{{\rm L,R}}$ are constant values
of the potential and scattering length, to the left and right of $x=L$,
where respective steps take place.

Notice that such potential steps may be realized in present BEC experiments
upon employing a detuned laser beam shined over a razor edge to make a sharp barrier,
with the diffraction-limited fall-off of the laser intensity being smaller
than the healing length of the condensate; in such a situation, the potential
can be effectively described by a step function.
On the other hand, the implementation of nonlinearity steps can be based on
the interaction tunability of specific atomic species by applying external magnetic
or optical fields \cite{FRM,FRM2}. For instance, confining ultracold atoms
in an elongated trapping potential near the surface of an atom chip \cite{chip}
allows for appropriate local engineering of the scattering length to form steps
(of varying widths), where the atom-surface separation sets a scale for
achievable minimum step widths. The trapping potential can
be formed optically, possibly also by a suitable combination
of optical and magnetic fields (see Ref.~\cite{lnl2} for a relevant discussion).

Measuring the longitudinal coordinate $x$ in units of $\sqrt{2}\xi$
(where $\xi \equiv \hbar /\sqrt{2m n g_{1D}}$ is the healing length),
time $t$ in units of $\sqrt{2} \xi/c_{s}$
(where $c_{s}\equiv \sqrt{g_{1D} n/m}$ is the speed of sound and $n$ is the peak density),
and energy in units of $g_{1D}n$, we cast Eq.~(\ref{gpe1}) to the
following dimensionless form (see Ref.~\cite{box_rein}):
\begin{eqnarray}
i\frac{\partial u}{\partial t} &=& - \frac{1}{2}\frac{\partial^2 u}{\partial x^2}
+ \frac{ \alpha(x)}{\alpha_{\rm{L}}} |u|^2  u + V(x) u,
\label{u}
\end{eqnarray}
where $u=\sqrt{n}\Psi$. Unless stated otherwise, in the simulations below
we fix the parameter values as follows:
$V_{0}=10$ and $w=250$ (for the box potential), $V_{\rm{L}}=0$ and
$V_{\rm{R}}=\pm 0.01$ for the potential step, as well as
$\alpha_{\rm L}=1$ and $a_{{\rm R}} \in [0.9,~1.1]$ for the nonlinearity step.
Nevertheless, our theoretical approach is general
(and will be kept as such in the exposition that follows in this
section).

Here we should mention that Eq.~(\ref{u}) can also be applied in the context
of nonlinear optics \cite{molnew}: in this case, $u$ represents the complex electric field
envelope, $t$ is the propagation distance and $x$ is the transverse direction, while
$V(x)$ and $\alpha(x)$ describe the (transverse) spatial profile of the linear and nonlinear
parts of the refractive index~\cite{kominis}.
This way, Eq.~(\ref{u}) can be used for the study of
optical beams, carrying dark solitons, near interfaces separating different optical media,
with (different) defocusing Kerr nonlinearities.

\subsection{Perturbation theory and equivalent particle picture}

Assuming that, to a first approximation, the box potential can be neglected, we consider the dynamics of
a dark soliton, which is located in the region $x<L$, and moves to the right, towards the potential
and nonlinearity steps (similar considerations for a soliton located in the region $x>L$
and moving to the left are straightforward).
In such a case, we seek for a solution of Eq.~(\ref{u}) in the form:
\begin{eqnarray}
u(x,t) &=& \sqrt{\mu_{\rm{L}} - V_{\rm{L}}}\exp{(-i \mu_{\rm{L}}t)} \upsilon(x,t),
\label{ic}
\end{eqnarray}
where $\mu_{\rm{L}}$ is the chemical potential, and the $\upsilon(x,t)$ is the wavefunction of the dark soliton.
Then, introducing the transformations $t \rightarrow \left( {\mu_{\rm{L}} - V_{\rm{L}}}\right) t$
and  $x \rightarrow  \sqrt{\mu_{\rm{L}}-V_{\rm{L}}} x$, we express Eq.~(\ref{u}) as a perturbed NLS
equation for the dark soliton:
\begin{eqnarray}
i\frac{\partial \upsilon}{\partial t}+\frac{1}{2}\frac{\partial^2 \upsilon}{\partial x^2}-\left( |\upsilon|^2-1 \right) \upsilon &=&  P(\upsilon).
\label{upsilon}
\end{eqnarray}
Here, the functional perturbation $P(\upsilon)$ has the form:
\begin{eqnarray}
P(\upsilon) &=&  \left(A+ B |\upsilon| ^2\right)  \upsilon \mathcal{H}(x-L),
\label{P}
\end{eqnarray}
where $\mathcal{H}$ is the Heaviside step function, and coefficients $A$,~$B$ are given by:
\begin{eqnarray}
A &=& \frac{V_{\rm{R}}-V_{\rm{L}}}{\mu_{\rm{L}}-V_{\rm{L}}}, \quad
B  = \frac{ \alpha_{\rm{R}}}{\alpha_{\rm{L}}}-1.
\end{eqnarray}
These coefficients, which set the magnitudes of the potential and nonlinearity steps,
are assumed to be small. Such a situation corresponds, e.g., to the case where
$\mu_{\rm{L}}=1$, $V_{\rm{L}}=0$, $V_{\rm{R}} \sim \epsilon$, and
$a_{\rm{R}}/\alpha_{\rm{L}} \sim 1$, where $0<\epsilon\ll 1$
is a formal small parameter (this choice will be used in our simulations below).
In the present work, we assume that the jump from left to right is
``sharp'', i.e., we do not explore the additional possibility of a
finite width interface. If such a finite width was present but was
the same between the linear and nonlinear interface, essentially
the formulation below would still be applicable, with the
Heaviside function above substituted by a suitable smoothened
variant (e.g. a $\tanh$ functional form). A more complicated
setting deferred for future studies would involve the existence
of two separate widths in the linear and nonlinear step and the
length scale competition that that could involve.

Equation~(\ref{upsilon}) can be studied analytically upon employing
perturbation theory for dark solitons (see, e.g., Refs.~\cite{yskyang,djf,MJA}):
first we note that, in the absence of the perturbation $(\ref{P})$,
Eq.~(\ref{upsilon}) has a dark soliton solution of the form:
\begin{eqnarray}
\upsilon(x,t) = \cos\phi \tanh X +i\sin\phi,
\label{ds}
\end{eqnarray}
where $X = \cos\phi[x - x_0(t)]$ is the soliton coordinate, $\phi$ is the soliton phase angle
$(|\phi|<\pi/2)$ describing the darkness of the soliton,
$\cos\phi$ is the soliton depth ($\phi=0$ and $\phi \ne 0$ correspond to
stationary black solitons and gray solitons, respectively), while $x_0(t)$
and $dx_0/dt = \sin\phi$ denote the soliton center and velocity, respectively.
Then, considering an adiabatic evolution of the dark soliton, we assume that in the presence
of the perturbation the dark soliton parameters become slowly-varying unknown
functions of time $t$. Thus, the soliton phase angle becomes $\phi\rightarrow \phi(t)$ and,
as a result, the soliton coordinate becomes $X=\cos\phi(t)\big(x-x_0(t)\big)$,
with $dx_0(t)/dt=\sin\phi(t)$.

The evolution of the soliton phase angle can be found by means of the evolution of the
renormalized soliton energy, $E_{ds}$, given by \cite{yskyang,djf}:
\begin{eqnarray}
E_{ds} &=& \frac{1}{2} \int ^{\infty}_{-\infty}{\Big[|\upsilon_x|^2
+\left( |\upsilon|^2-1\right)^2\Big] dx}.
\label{E_L}
\end{eqnarray}
Employing Eq.~(\ref{ds}), it can readily be found that $dE_{ds}/dt=-4 \cos^2\phi~\sin\phi~d\phi / dt$.
On the other hand, using Eq.~(\ref{upsilon}) and its complex conjugate, yields
the evolution of the renormalized soliton energy:
$dE_{ds} / dt = - \int ^{+\infty}_{-\infty} \big(P \bar{\upsilon}_t + \bar{P} \upsilon_t \big) dx$,
where bar denotes complex conjugate.
Then, the above expressions for $dE_{ds}/dt$ yield the evolution of $\phi$, namely
\begin{eqnarray}
\frac{d\phi}{dt} &=& \frac{1}{2\cos^2\phi \sin\phi} {\rm Re} \Big\{\int ^{+\infty}_{-\infty}
P(\upsilon) \bar{\upsilon}_t dx\Big\}.
\label{dphi1}
\end{eqnarray}
Inserting the perturbation (\ref{P}) into Eq.~(\ref{dphi1}), and
performing the integration, we obtain the following result:
\begin{eqnarray}
\frac{d\phi}{dt} =&-&\frac{1}{4}\big( A+ B\big) \sech^2 \big(L-x_0\big) \nonumber \\
&+& \frac{1}{8} B \sech^4 \big(L-x_0\big),
\label{dphi2}
\end{eqnarray}
where we have considered the case of nearly stationary (black) solitons with $\cos\phi \thickapprox 1$
(and $\sin\phi \thickapprox \phi$). Combining Eq.~(\ref{dphi2}) with the
above mentioned equation for the soliton velocity, $dx_0(t)/dt=\sin\phi(t)$,
we can readily derive the following equation for motion for the soliton center:
\begin{eqnarray}
\frac{d^2 x_0}{dt^2} = - \frac{dW}{d x_0},
\label{eqmot}
\label{eqmotion}
\end{eqnarray}
where the effective potential $W(x_0)$ is given by:
\begin{eqnarray}
W(x_0)=&-&\frac{1}{8}\big( 2A + B \big) \tanh\big(L-x_0\big) \nonumber \\
&-&\frac{1}{24} B \tanh^3\big(L-x_0\big).
\label{W}
\end{eqnarray}

\subsection{Forms of the effective potential}

The form of the effective potential suggests that fixed points, where -- potentially -- dark solitons
may be trapped, exist only in the presence of the nonlinearity step ($B\ne 0$).
I.e., it is the competition between the linear and nonlinear step that
enable the presence of fixed points and associated more complex dynamics;
in the presence of solely a linear step, the dark soliton encounters
solely a step potential, similarly to what is the case for its bright
sibling~\cite{nogami}; see also below.

\begin{figure}[tbp]
\centering
\includegraphics[scale=0.38]{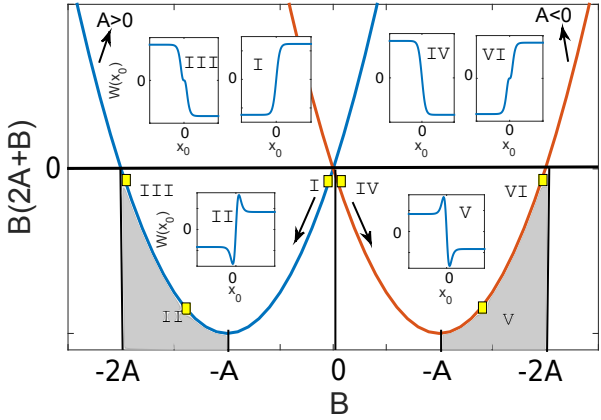}
\caption{Sketch showing domains of existence of fixed points of the effective potential $W(x_0)$
(depicted by gray areas) for $A>0$ (blue line) and $A<0$ (red line). The insets ${\rm I}-{\rm III}$
(${\rm IV}-{\rm VI}$) show the form of $W(x_0)$, starting from -- and ending to -- a small
finite value of nonlinearity step $B$, which is gradually decreased (increased)
for $A>0$ ($A<0$), cf. black arrows. Small rectangular (yellow) points indicate
parameter values corresponding to the forms of $W(x_0)$ shown in the
insets ${\rm I}-{\rm VI}$.
}
\label{exis}
\end{figure}

In fact, in our setting it is
straightforward to find that there exist two fixed points, located at:
\begin{eqnarray}
x_{0\pm} &=& \frac{1}{2} \ln \left(\frac{-A\mp \sqrt{-B\left(2A+B\right)}}{A+B}\right),
\label{fp}
\end{eqnarray}
for $B(2A+B)<0$ and $-2A<B<-A$, for $A>0$, or $-A<B<-2A$, for $A<0$.
In Fig.~\ref{exis} we plot $B(2A+B)$ as a function of $B$, for $A>0$ (blue line) and $A<0$ (red line).
The corresponding domains of existence of the fixed points, are also depicted by the gray areas.
Insets show typical profiles of the effective potential $W(x_0)$, for different values of $B$,
which we discuss in more detail below. From the figure (as well as
from Eq.~(\ref{fp}) itself), the saddle-center nature of the bifurcation
of the two fixed points, which are generated concurrently ``out of the blue
sky'' is immediately evident.

First, we consider the case of the absence of the nonlinearity step, $B=0$,
as shown in the insets ${\rm I}$ and ${\rm IV}$ of Fig.~\ref{exis}, for $A>0$ and $A<0$, respectively.
In this case, $W(x_0)$ assumes a step profile, induced by the potential step. This form is preserved
in the presence of a finite nonlinearity step, $B\ne 0$, namely for $-A<B<0$ and $0<B<-A$, in the cases
$A>0$ and $A<0$, respectively.

A more interesting situation occurs when the nonlinearity step further decreases (increases), and takes
values $-2A<B<-A$ for $A>0$, or $-A<B<-2A$ for $A<0$. In this case, the effective potential
features a local minimum (maximum), i.e., an elliptic (hyperbolic) fixed point,
in the region $x<0$ ($x>0$) for $A>0$ emerge (as per the
saddle-center bifurcation mentioned above) close to the location of
the potential and nonlinearity steps, i.e., near $x=0$; a similar situation occurs for $A<0$, but the
local minimum becoming a local maximum, and vice versa. The locations $x_{0\pm}$ of the fixed points are
given by Eq.~(\ref{fp}); as an example, using parameter values $V_{\rm{L}}=0$, $V_{\rm R}=-0.01$,
$\alpha_{\rm L}=1$ and $\alpha_{\rm R}=1.015$, we find that $x_{0+}=0.66$ ($x_{0-}=-0.66$)
for the elliptic (hyperbolic) fixed point.

As the nonlinearity step becomes deeper, the asymptotes (for $x \rightarrow \pm \infty$) of $W(x_0)$
become smaller and eventually vanish. For fixed $V_{\rm L}=0$ (and $\mu_{\rm L}=1$),
Eq.~(\ref{W}) shows that this happens for $B=-(3/2)A$; in this case,
the potential features a ``spiky'' profile,
in the vicinity of $x=0$ (see, e.g., upper panel of Fig.~\ref{pp1} below).
For $B<-(3/2)A$, the asymptotes of $W(x_0)$ become finite again, and take
a positive (negative) value for $x<0$, and a negative (positive) value for $x>0$, in the case $A>0$ ($A<0$).
The spiky profile of $W(x_0)$ in the vicinity of $x=0$ is preserved in this case too, but as
$B$ decreases it eventually disappears, as shown
in the insets ${\rm III}$ and ${\rm VI}$ of Fig.~\ref{exis}.

\subsection{Solitons at the fixed points of the effective potential}

\begin{figure}[tbp]
\centering
\includegraphics[scale=0.3]{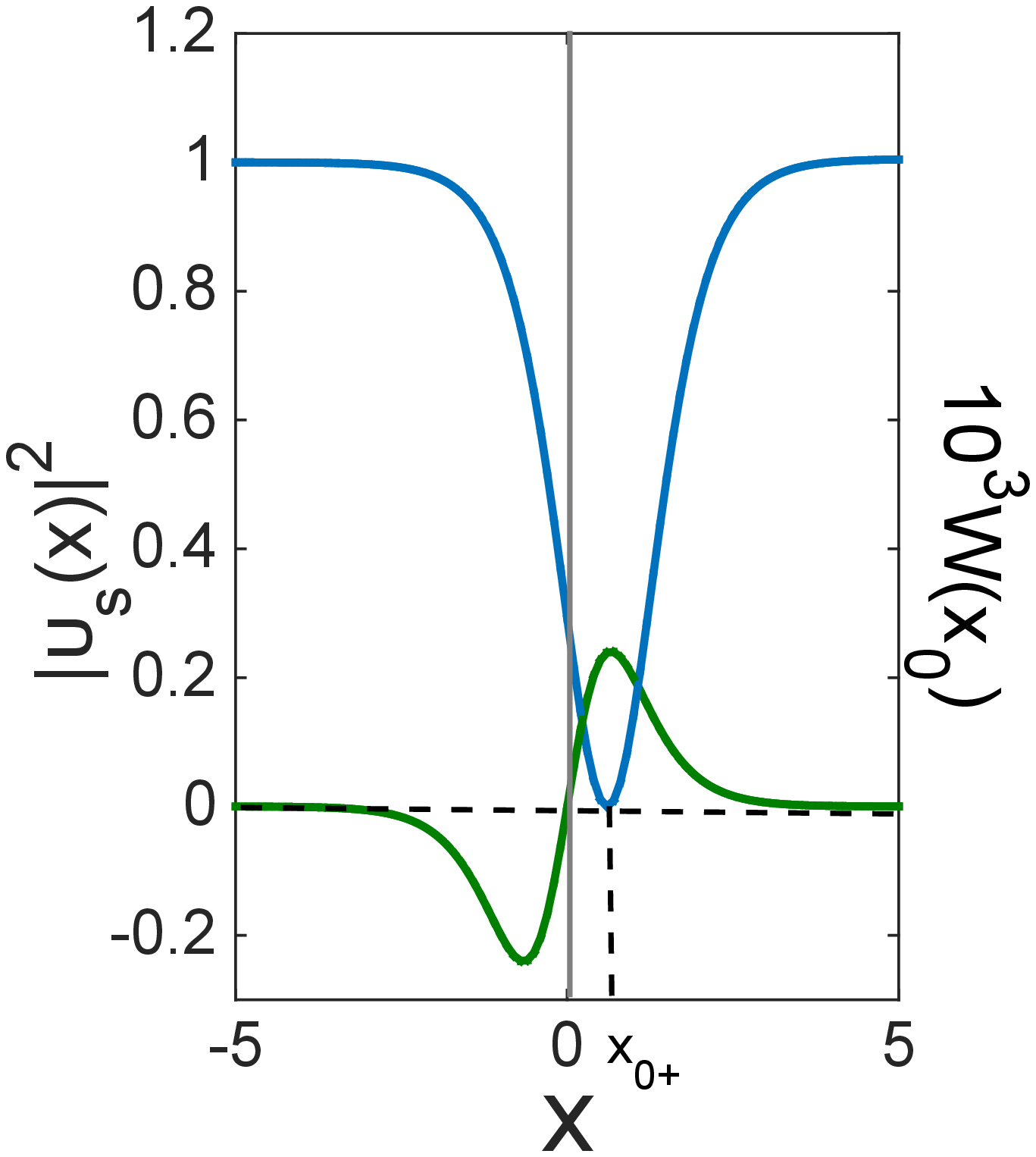}
\includegraphics[scale=0.3]{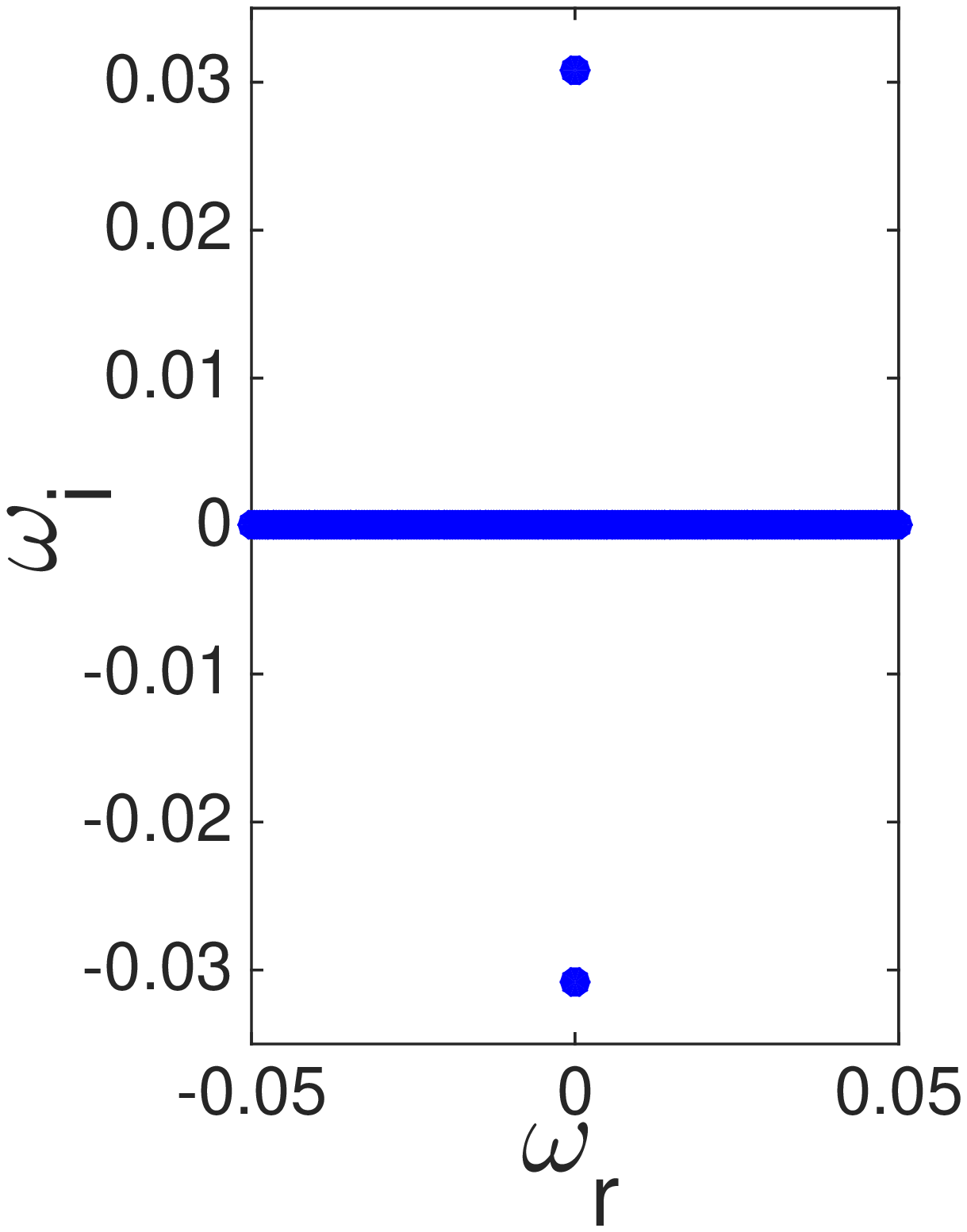}
\caption{(Color online) Left panel: density profile of the stationary soliton (blue line) at
the hyperbolic fixed point $x_{0+}=0.66$, as found
numerically, using the ansatz $\upsilon_s(x)=[1-V(x)]^{1/2}\tanh(x)$ in Eq.~(\ref{ss}),
for $\alpha_{\rm R}/\alpha_{\rm L}=0.985$, $V_{\rm R}=0.01$, $V_{\rm{L}}=0$, $\mu_{\rm{L}}=1$; green line illustrates the
corresponding effective potential $W(x_0)$. Right panel: corresponding spectral plane
($\omega_r$,~$\omega_i$) of the corresponding eigenfrequencies, illustrating an
exponential growth due to an imaginary eigenfrequency pair.
}
\label{bdg1}
\end{figure}
\begin{figure}[tbp]
\centering
\includegraphics[scale=0.3]{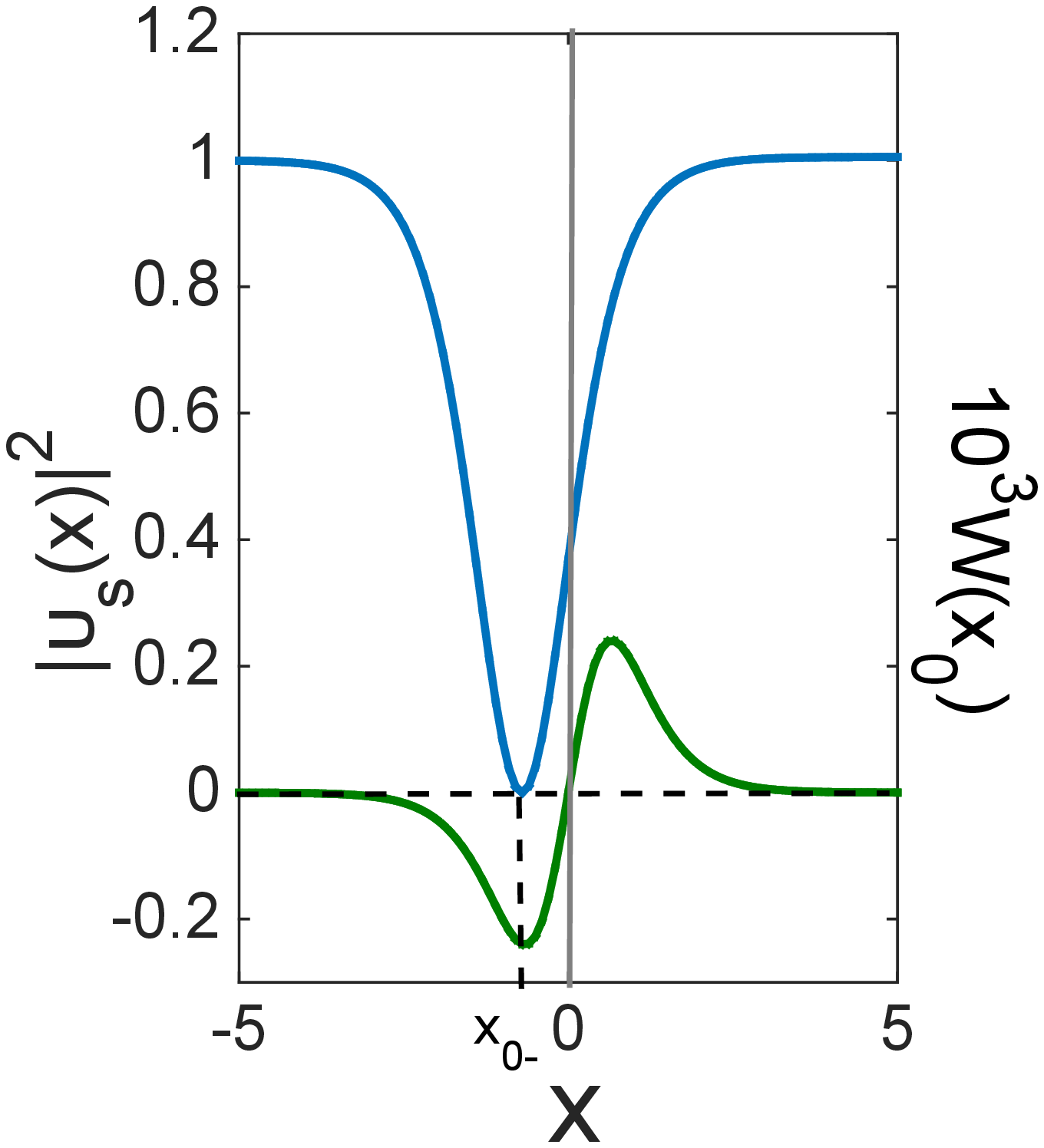}
\includegraphics[scale=0.3]{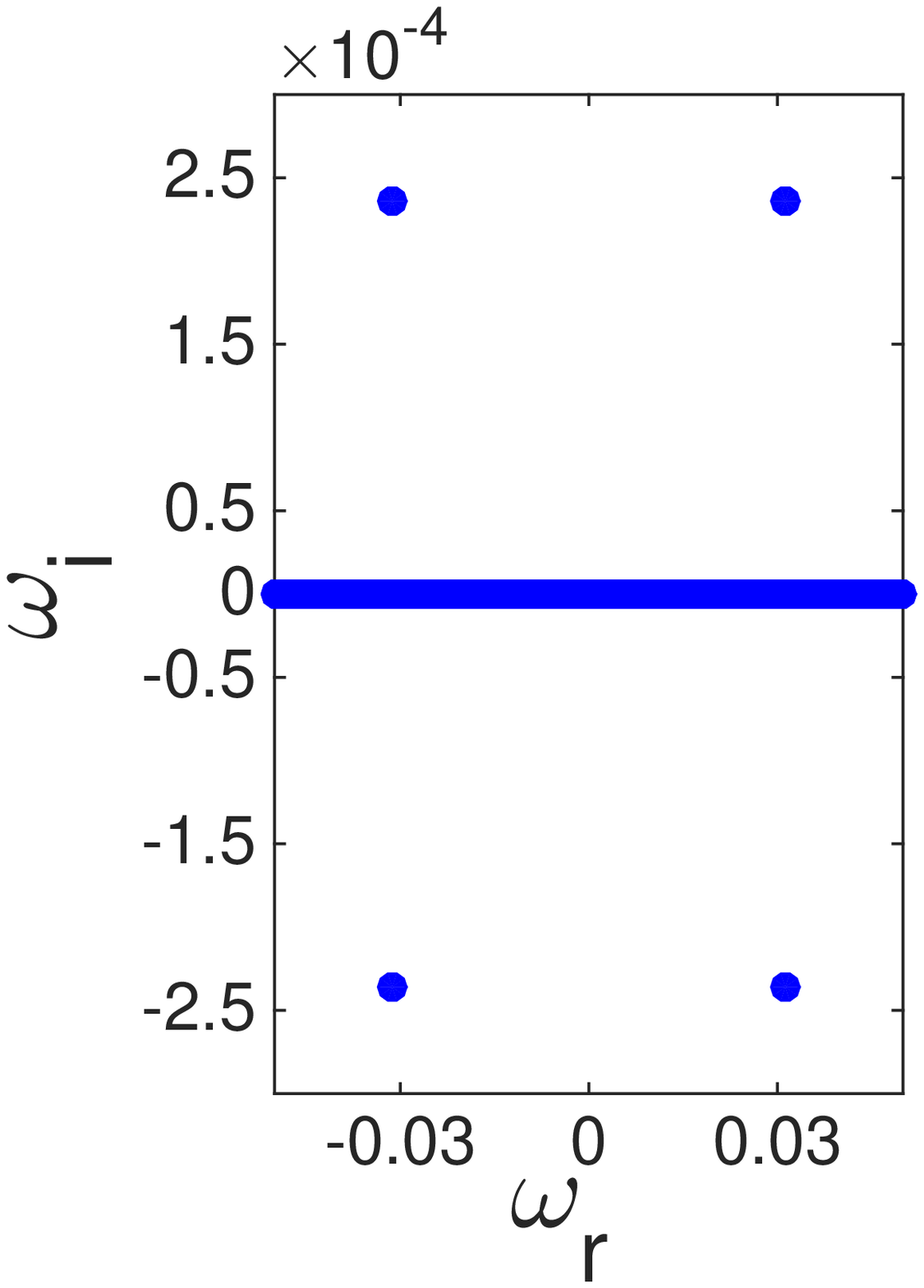}
\caption{(Color online) Same as Fig.~\ref{bdg1}, but for a soliton located
at the elliptic fixed point $x_{0-}=-0.66$; this state is found using
the initial ansatz
$\upsilon_s(x)=[1-V(x)]^{1/2}\tanh(x+0.2)$.
The spectral plane in the right panel illustrates an
oscillatory growth due to the presence of a complex quartet of eigenfrequencies.
}
\label{bdg2}
\end{figure}

The above analysis poses an interesting question regarding the existence of
stationary solitons of Eq.~(\ref{u}) at the fixed points of the effective potential.
To address this question,
we use the ansatz $u(x,t)=\exp(-it)\upsilon_s(x)$, for a stationary soliton $\upsilon_s(x)$, and
obtain from Eq.~(\ref{u}) the equation:
\begin{eqnarray}
\upsilon_s = -\frac{1}{2}\frac{d^2\upsilon_s}{dx^2}+\frac{\alpha(x)}{\alpha_{\rm L}}|\upsilon_s|^2 \upsilon_s
+V(x)\upsilon_s.
\label{ss}
\end{eqnarray}
Notice that we have assumed without loss of generality a unit
frequency solution; the formulation below can be used at will for
any other frequency.
The above equation is then solved numerically, by means of Newton's method, employing the ansatz (for $L=0$):
\begin{equation}
\upsilon_s(x)=[1-V(x)]^{1/2}\tanh(x-x_{0}).
\label{anstz}
\end{equation}
As shown in the left panel of Fig.~\ref{bdg1}, assuming an ansatz
within Eq.~(\ref{anstz}) in which the soliton is initially
placed at $x_0=0$, we find a steady state exactly at the hyperbolic fixed point $x_{0+}=0.66$, as
found from Eq.~(\ref{fp}). On the other hand, the left panel of Fig.~\ref{bdg2} shows a case
where the initial guess is assumed in Eq.~(\ref{anstz})
to have a soliton positioned at $x_0=-0.2$,
which leads to a stationary soliton located exactly at the elliptic fixed point $x_{0-}=-0.66$
predicted by Eq.~(\ref{fp}).

It is now relevant to study the stability of these stationary soliton states,
performing a Bogoliubov-de Gennes (BdG) analysis \cite{pita,book_fr,djf}.
We thus consider small perturbations of $\upsilon_s(x)$, and seek
solutions of Eq.~(\ref{ss}) of the form:
\begin{eqnarray}
u(x,t) =e^{-it} \left[\upsilon_s(x)
+\delta \left(b(x)e^{-i\omega t} +\bar{c}(x) e^{i\bar{\omega} t}\right)\right],
\label{ban}
\end{eqnarray}
where $(b(x),~c(x))$ are eigenmodes, $\omega=\omega_r +i \omega_i$ are
(generally complex) eigenfrequencies, and $\delta \ll 1$.
Notice that the occurrence of a complex eigenfrequency always leads to a dynamic instability; thus, a
linearly stable configuration is tantamount to $\omega_i = 0$ (i.e., all eigenfrequencies are real).

Substituting Eq.~(\ref{ban}) into Eq.~(\ref{u}), and linearizing with respect to $\delta$, we derive
the following BdG equations:
\begin{eqnarray}
&&\left[\hat{H} -1+ 2\frac{\alpha(x)}{\alpha_{\rm L}}\upsilon_s^2 \right] b
+ \frac{\alpha(x)}{\alpha_{\rm L}} \upsilon_s^2 c
=\omega b, \\
&&\left[\hat{H}-1+ 2\frac{\alpha(x)}{\alpha_{\rm L}}\upsilon_s^2 \right]c
+ \frac{\alpha(x)}{\alpha_{\rm L}} \upsilon_s^{2} b=-\omega c,
\label{matrices}
\end{eqnarray}
where $\hat{H}=-(1/2)\partial_{x}^2+V(x)$ is the single particle operator. This eigenvalue
problem is then solved numerically. Examples of the stationary dark solitons at
the fixed points $x_{0\pm}$ of the effective potential $W$, as well as their corresponding
BdG spectra, are shown in Figs.~\ref{bdg1} and \ref{bdg2}.
It is observed that the solitons are dynamically unstable, as seen by the presence of
eigenfrequencies with nonzero imaginary part in the spectra, although
the mechanisms of instability are distinctly different between
the two cases (of Figs.~\ref{bdg1} and \ref{bdg2}).

\begin{figure}[tbp]
\centering
\includegraphics[scale=0.3]{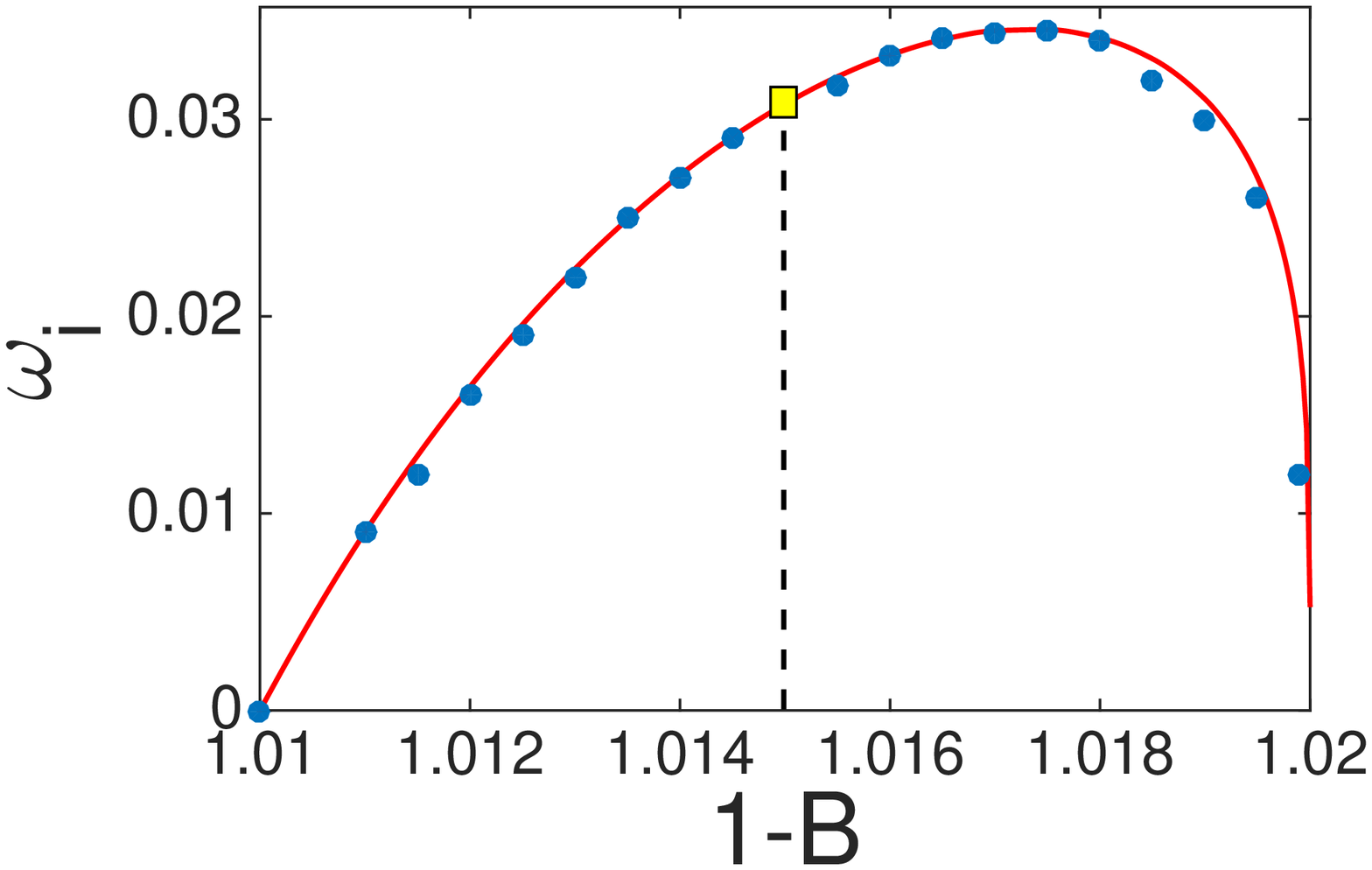}
\includegraphics[scale=0.3]{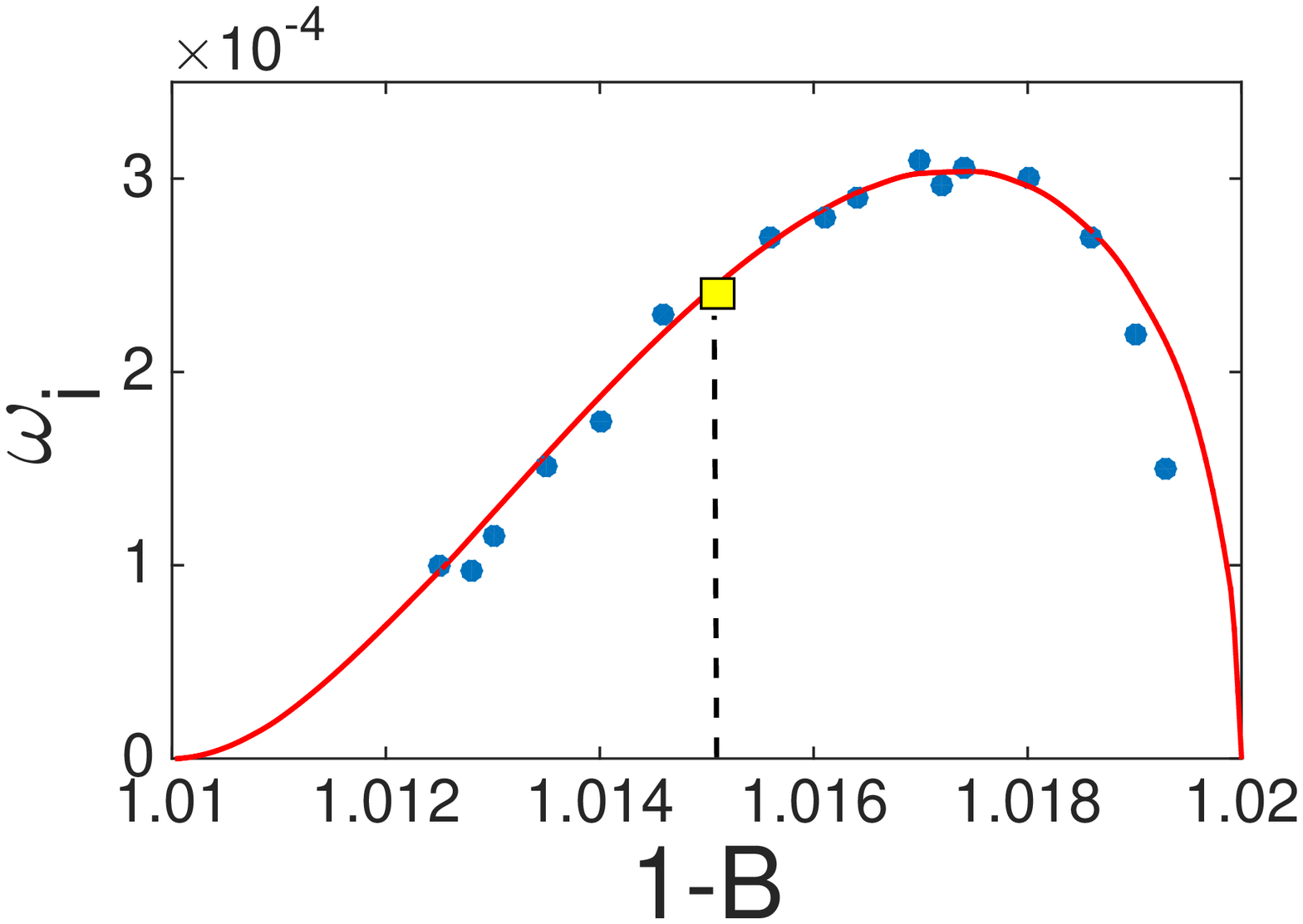}
\includegraphics[scale=0.33]{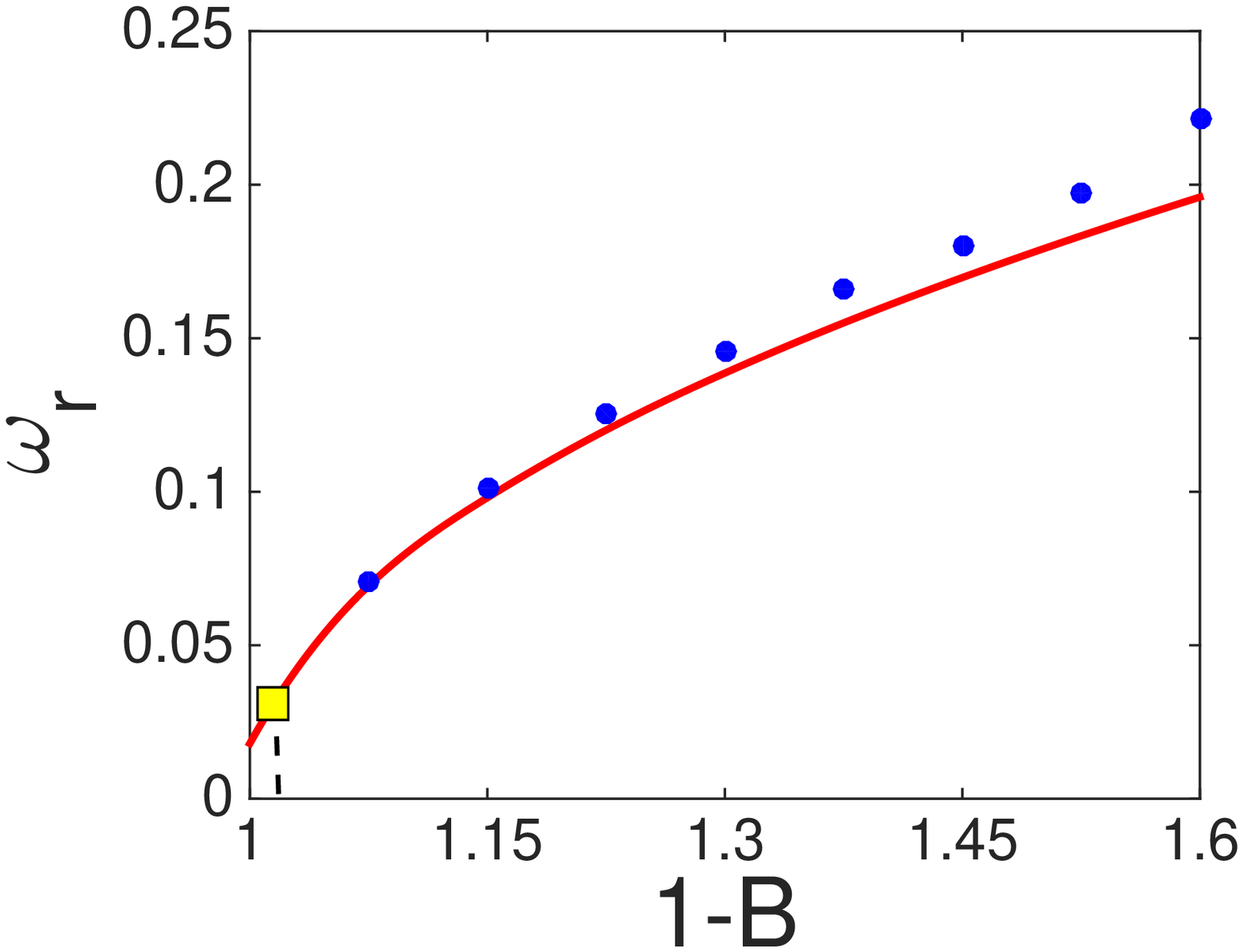}
\caption{(Color online) Top panel: the imaginary part of the eigenfrequency, $\omega_i$,
as a function of $1-B$ (with $B<0$), for a soliton located at the hyperbolic fixed point,
$x=x_{0+}$. Middle and bottom panels show the dependence of imaginary and real parts,
$\omega_i$ and $\omega_r$, of the eigenfrequency on $1-B$, for a soliton located at
the elliptic fixed point, $x=x_{0-}$, i.e., the case that leads
to an eigenfrequency quartet. Solid blue curves correspond to the analytical prediction
[cf. Eqs.~(\ref{ch_eq}) and~(\ref{M''})], blue circles depict numerical results, while yellow
squares depict eigenfrequency values corresponding to the cases shown in Figs.~\ref{bdg1}
and~\ref{bdg2}. For the top and middle panels $A=0.01$, while for the bottom panel $A=-(2/3)B$; 
in all cases, $\mu_{\rm{L}}=1$.}
\label{continuation}
\end{figure}

To better understand these instabilities, and also provide an analytical estimate for the relevant
eigenfrequencies, we may follow the analysis of Ref.~\cite{zamp}; see also Ref.~\cite{zamp_fr} for
application of this theory to the case of a periodic, piecewise-constant scattering length setting.
According to these works, solitons persist in the presence of the perturbation $P(\upsilon)$
of Eq.~(\ref{P}) (of strength $A,~B\sim \epsilon$) provided that the
Melnikov function condition
\begin{eqnarray}
M'(x_0) = \int^{+\infty}_{-\infty} \frac{\partial P(\upsilon)}{\partial x}{\rm sech}^2(x-x_0) dx = 0,
\label{M'}
\end{eqnarray}
possesses at least one root, say $\tilde{x}_0$.
Then, the stability of the dark soliton solutions at $x_{0\pm}$
depends on the sign of the derivative of the function in Eq.~(\ref{M'}), evaluated at
$\tilde{x}_0$: an instability occurs, with one imaginary eigenfrequency pair for
$\epsilon M''(\tilde{x}_0)<0$, and with exactly one complex eigenfrequency quartet for
$\epsilon M''(\tilde{x}_0)>0$. The instability is dictated by the translational eigenvalue,
which bifurcates from the origin as soon as the perturbation is present.
For $\epsilon M''(\tilde{x}_0)<0$, the relevant eigenfrequency pair moves
along the imaginary axis, leading to an immediate instability associated
with exponential growth of a perturbation along the relevant eigendirection.
On the other hand, for $\epsilon M''(\tilde{x}_0)>0$, the eigenfrequency moves along the real
axis; then, upon collision with eigenfrequencies of modes of opposite signature
than that of the translation mode,
it gives rise to
a complex eigenfrequency quartet, signaling the presence of an oscillatory instability.
The relevant eigenfrequencies can be determined by a
quadratic characteristic equation which takes the form \cite{zamp},
\begin{eqnarray}
\lambda^2+\frac{1}{4}M''(\tilde{x}_0)\left(1-\frac{\lambda}{2}\right) = \mathcal{O}(\epsilon^2),
\label{ch_eq}
\end{eqnarray}
where eigenvalues $\lambda$ are related to eigenfrequencies $\omega$ through $\lambda^2=-\omega^2$.
Since the roots of $M''(x_0)$ are the two fixed points $x_{0\pm}$, we may evaluate
$M''(x_{0\pm})$ explicitly, and obtain:
\begin{eqnarray}
M''(x_{0\pm})=&-&2{\rm sech}^2(x_{0\pm})\tanh(x_{0\pm}) \nonumber \\
&\times&\left[A+B\tanh^2(x_{0\pm}) \right].
\label{M''}
\end{eqnarray}
To this end, combining Eqs.~(\ref{ch_eq}) and~(\ref{M''}) yields an analytical
prediction for the magnitudes of the relevant eigenfrequencies, for the cases of
solitons located at the hyperbolic or the elliptic fixed points of $W(x_0)$.

Figure~\ref{continuation} shows pertinent analytical results [depicted by (red) solid lines],
which are compared with corresponding numerical results [depicted by (blue) points].
In particular, the top panel of the figure illustrates the dependence of the
imaginary part of the eigenfrequency (real part of the
eigenvalue) $\omega_i$ on the parameter $1-B$ (with $B<0$),
for a soliton located at the hyperbolic fixed point, $x=x_{0+}$; this case
is associated with the scenario $M''(x_0)<0$. The middle and bottom panels of the figure
shows the dependence of $\omega_i$ and $\omega_r$ on $1-B$, but
for a soliton located at the elliptic fixed point, $x=x_{0-}$; in this case,
$M''(x_0)>0$, corresponding to an oscillatory instability as mentioned above.
It is readily observed that the agreement between the theoretical prediction
of Eqs.~(\ref{ch_eq}) and~(\ref{M''}) and the numerical result is very good; especially,
for values of $1-B$ close to unity, i.e., in the case $|B| \lesssim 0.15$ where perturbation theory
is more accurate, the agreement is excellent.

We should also remark here that a similarly good agreement between analytical and
numerical results was also found (results not shown here) upon
using as an independent parameter the strength of the potential step ($\sim A$),
instead of the strength of the nonlinearity step ($\sim B$), as in
the case of Fig.~\ref{continuation}.

\section{Dark solitons dynamics}

We now turn our attention to the dynamics of dark solitons near the potential and nonlinearity steps.
We will use, as a guideline, the analytical results presented in the previous section, and particularly the
form of the effective potential. Our aim is to study the scattering of a dark soliton, initially located in
the region $x<L$ and moving to the right, at the potential and nonlinearity steps (similar considerations,
for a soliton located in the region $x>L$ and moving to the left, are straightforward, hence only limited examples of the latter type will be presented).
We will consider the
scattering process in the presence of: (a) a single potential step, (b) a potential and nonlinearity step, and
(c) two potential and nonlinearity steps.

Attention will be paid to possible trapping of the soliton in the
vicinity of the location of the potential and nonlinearity steps,
and particularly at the hyperbolic fixed point (when present) of
the effective potential. Notice that in the context of optics such a soliton trapping effect
could be viewed as a formation of surface dark solitons at the interfaces between optical media
exhibiting different linear refractive indices and different defocusing Kerr nonlinearities (or atomic media bearing different linear potential and
interparticle
interaction properties at the two sides of the interface).

\subsection{A single potential step}

\begin{figure}[tbp]
\centering
\includegraphics[scale=0.38]{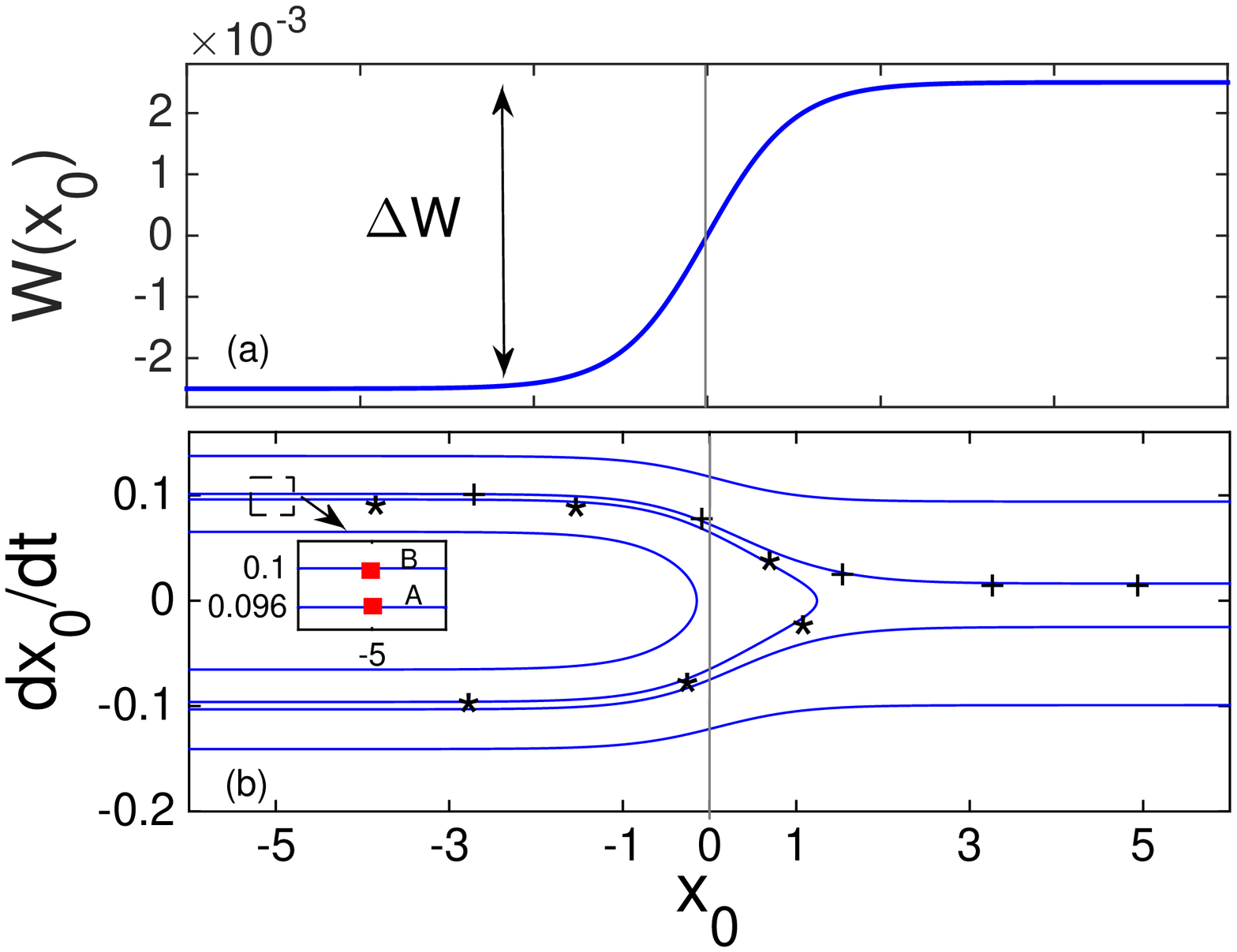}
\includegraphics[scale=0.195]{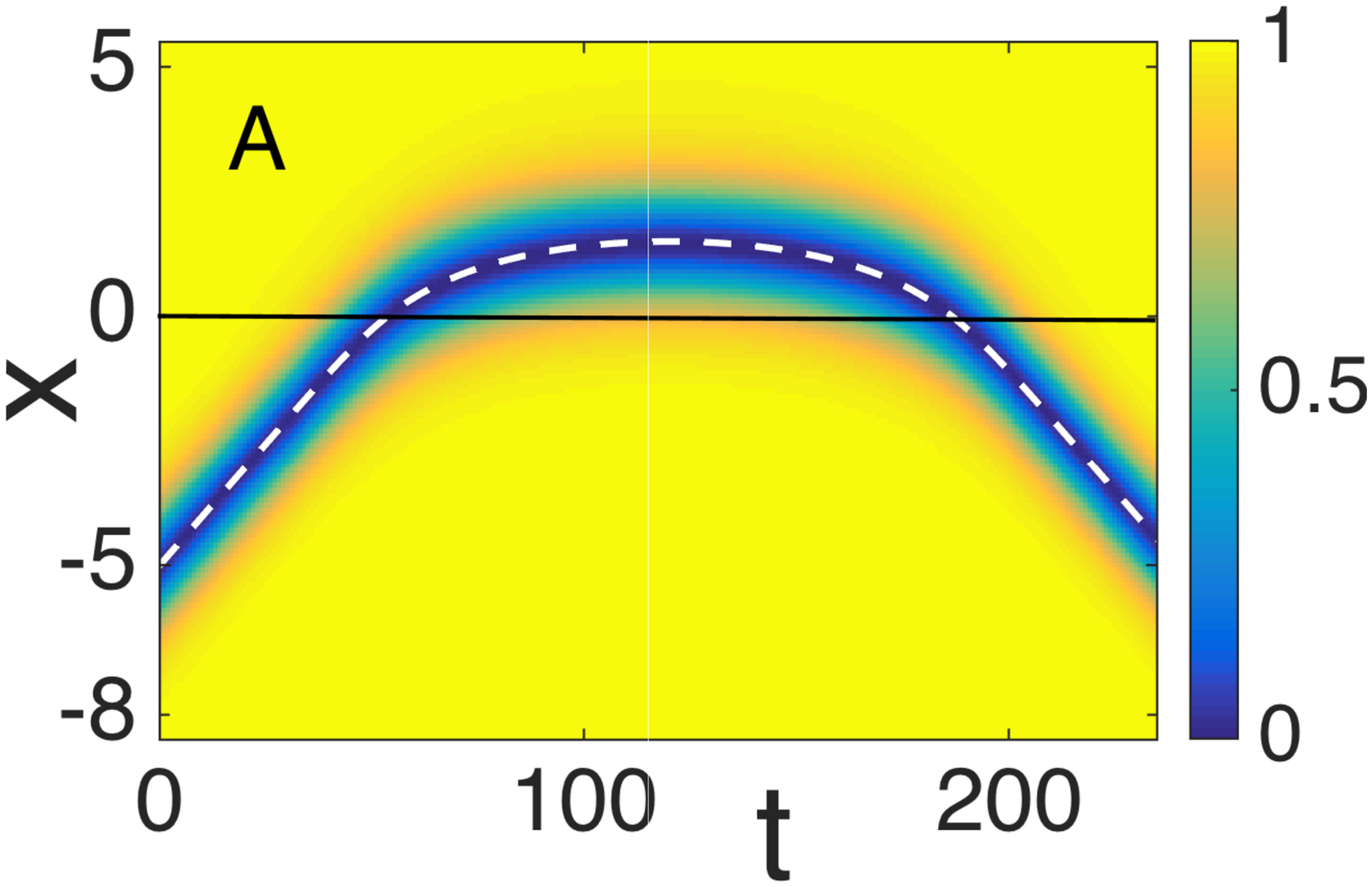}
\includegraphics[scale=0.195]{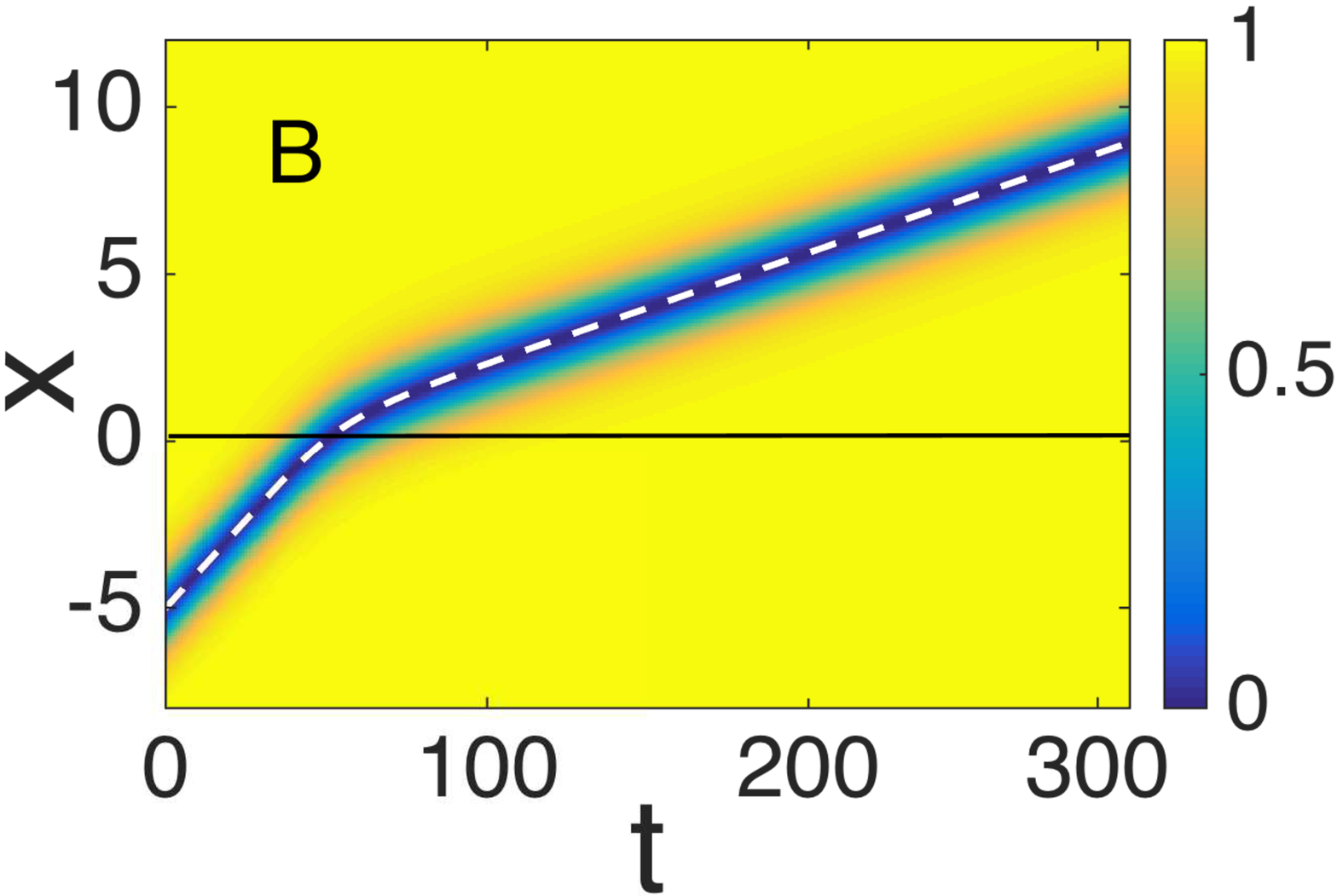}
\caption{(Color online) The case of a single potential step, $A=0.01$ and $B=0$,
corresponding to $V_{\rm{L}}=0$, $V_{\rm{R}}=0.01$,
$\alpha_{\rm{R}}=\alpha_{\rm{L}}$, and $\mu_{\rm{L}}=1$.
Top panel (a): effective potential $W(x_0)$; shown also is the potential difference
$\Delta W= W(+\infty)-W(-\infty)=4.99\times 10^{-3}$.
Middle panel (b): corresponding phase plane; inset shows the initial conditions (red squares A and B)
for the trajectories corresponding to reflection or transmission, while stars and crosses depict
respective PDE results.
Bottom panel: contour plots showing the evolution of the dark soliton density
for the initial conditions depicted in
the middle panel, i.e., $x_0=-5$ and $\phi=9.6\times 10^{-2}$ (left),
or $\phi=0.1$ (right); note that, here, $\phi_{\rm c}=0.099$.
Thick (blue) solid lines show PDE results, while dashed (white) lines depict ODE results.
}
\label{ref_tr}
\end{figure}

Our first scattering ``experiment'' refers to the case of a potential step only, corresponding to
$A>0$ and $B=0$ (cf. inset~I in Fig.~\ref{exis}). In this case, the effective potential
has typically the form shown in the top panel of Fig.~\ref{ref_tr}, while the associated phase-plane
is shown in the middle panel of the same figure. Clearly, according to the
particle picture for the soliton of the previous section, a dark soliton incident from the left
towards the potential step can either be reflected or transmitted: if the soliton has a velocity
$v=dx_0/dt$, and thus a kinetic energy
\begin{equation}
K = \frac{1}{2} v^2=\frac{1}{2}\sin^2\phi \approx  \frac{1}{2}\phi^2,
\label{Ep}
\end{equation}
smaller (greater) than the effective potential step $\Delta W= W(+\infty)-W(-\infty)$, as shown in the
top panel of Fig.~\ref{ref_tr}, then it will be reflected (transmitted).
Notice the approximation ($\sin \phi \approx \phi$) here which
is applicable for low speeds/kinetic energies.
This consideration leads to
$\phi < \phi_{\rm c}$ or $\phi > \phi_{\rm c}$ for reflection or transmission, where
the critical value $\phi_{\rm c}$ of the soliton phase angle is given by:
\begin{eqnarray}
\phi_{\rm c} &=& \sqrt{2 \Delta W}.
\label{kc}
\end{eqnarray}

In the numerical simulations, we found that the threshold between the two cases is
quite sharp and is accurately predicted by Eq.~(\ref{kc}). Indeed, consider the scenario shown in
Fig.~\ref{ref_tr}, corresponding to parameter values $V_{\rm{L}}=0$, $V_{\rm{R}}=0.01$,
$\alpha_{\rm{R}}=\alpha_{\rm{L}}$ and $\mu_{\rm{L}}=1$. In this case,
we find that $\Delta W=4.99\times 10^{-3}$,
which leads to the critical value (for reflection/transmission) of the soliton phase angle
$\phi_{\rm c}=9.99\times 10^{-2}$. Then, for a soliton initially placed at
$x_0=-5$, and for initial velocities corresponding to phase angles $\phi=9.6\times 10^{-2}$ or
$\phi=0.1$, we observe reflection or transmission, respectively. The corresponding soliton
trajectories are depicted both in the phase plane $(x_0, dx_0/dt)$ in the middle panel
of Fig.~\ref{ref_tr} and in the space-time contour plots showing the evolution of the
soliton density in the bottom panels of the same figure (see trajectories A and B for
reflection and transmission, respectively). Note that stars and crosses
in the middle panel correspond
to results obtained by direct numerical integration of the partial
differential equation (PDE), Eq.~(\ref{u}),
while the (white) dashed lines in the bottom panels depict results obtained by the
ordinary differential equation (ODE), Eq.~(\ref{eqmotion}). Obviously, the agreement between
theoretical predictions and numerical results is very good.

Here we should recall that in the case where the nonlinearity step is also present ($B\ne 0$), and
when $B>-A$ (for $A>0$) or $B<-A$ (for $A<0$), the form of the effective potential is similar
to the one shown in the top panel of Fig.~\ref{ref_tr}. In such cases, corresponding results
(not shown here) are qualitatively similar to the ones presented above (for $A\ne 0$ and $B=0$);
in addition, we have again captured accurately the velocity threshold for reflection/transmission.


\subsection{A potential and a nonlinearity step}

\begin{figure}[tbp]
\centering
\includegraphics[scale=0.38]{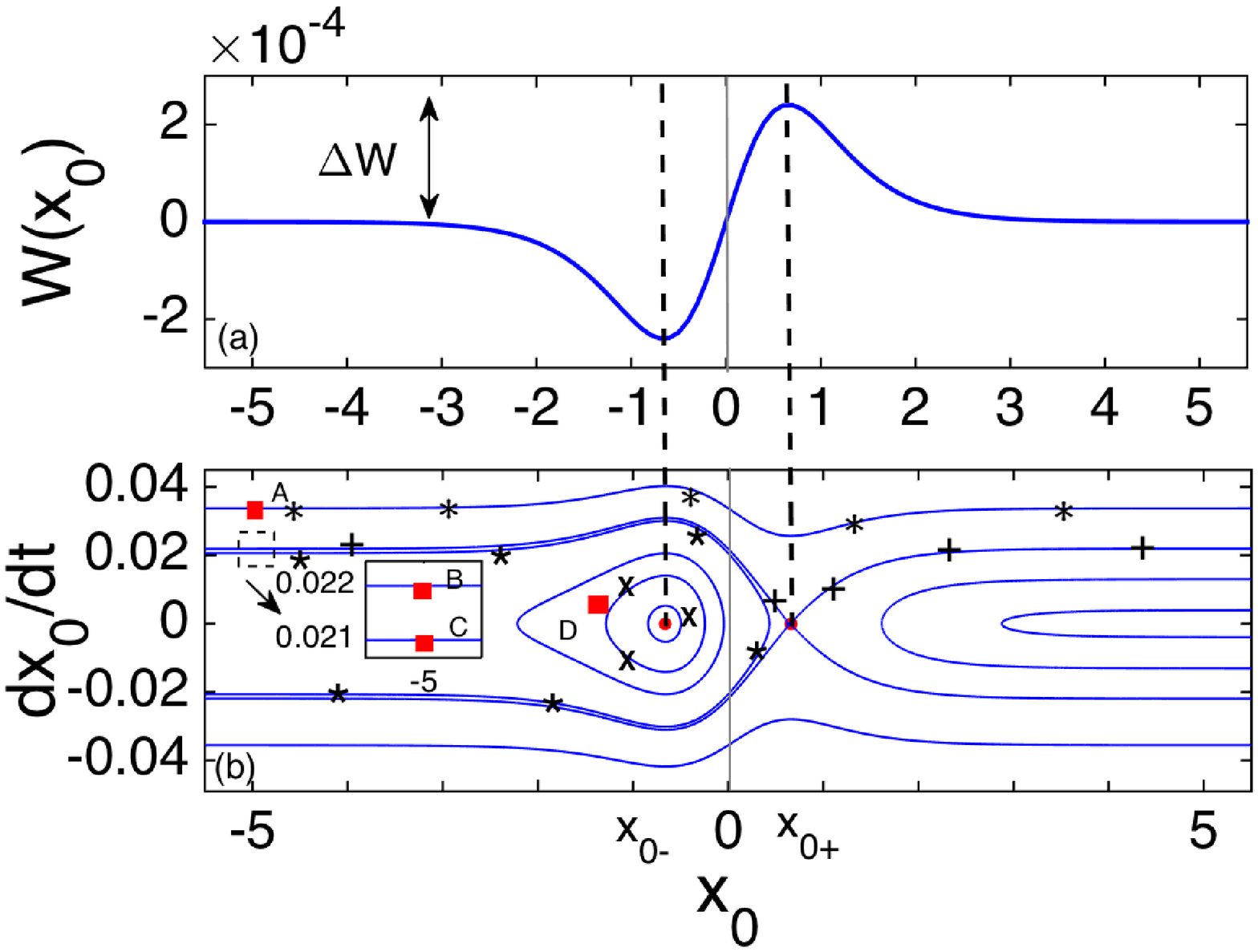}
\includegraphics[scale=0.195]{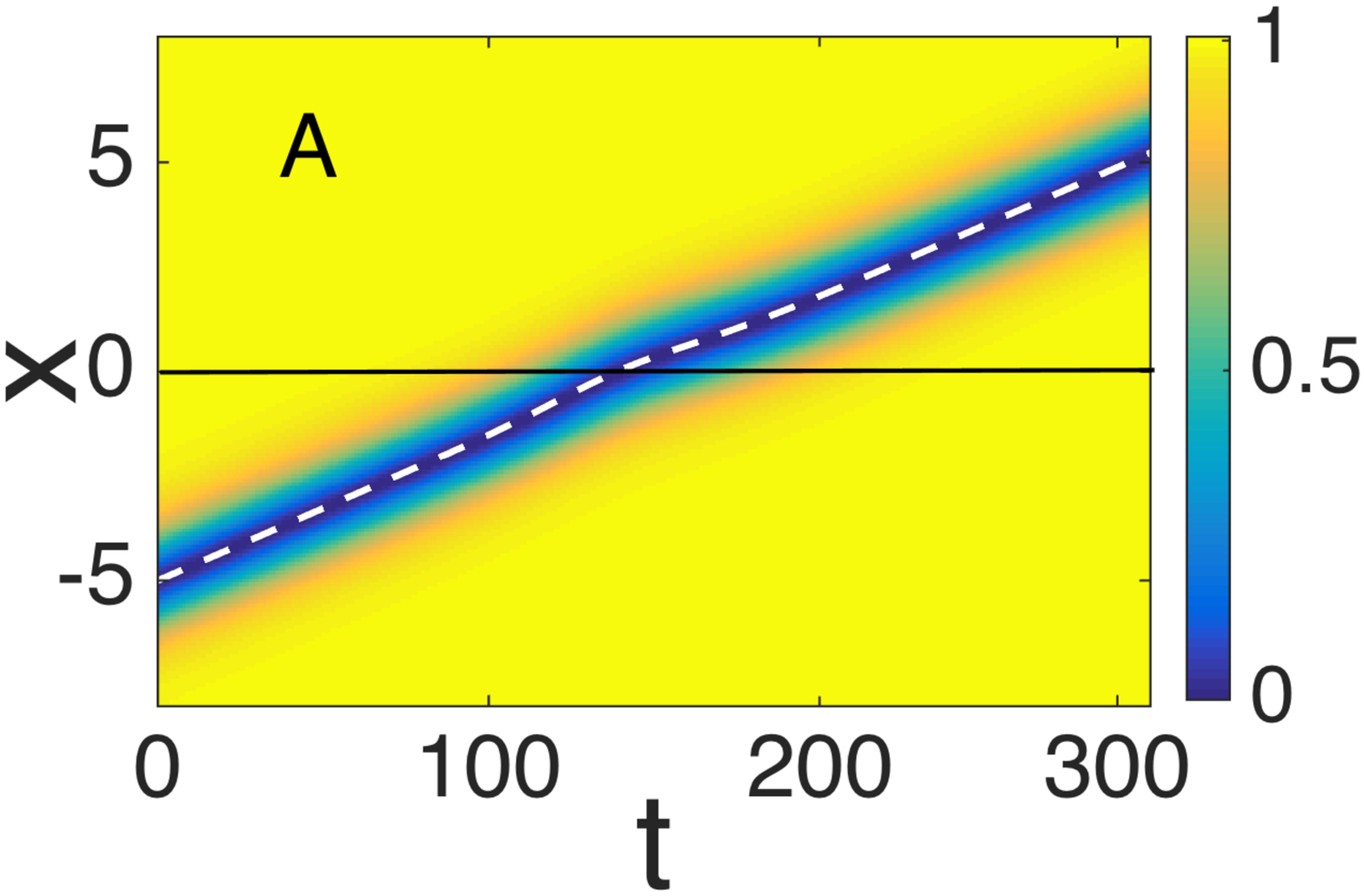}
\includegraphics[scale=0.195]{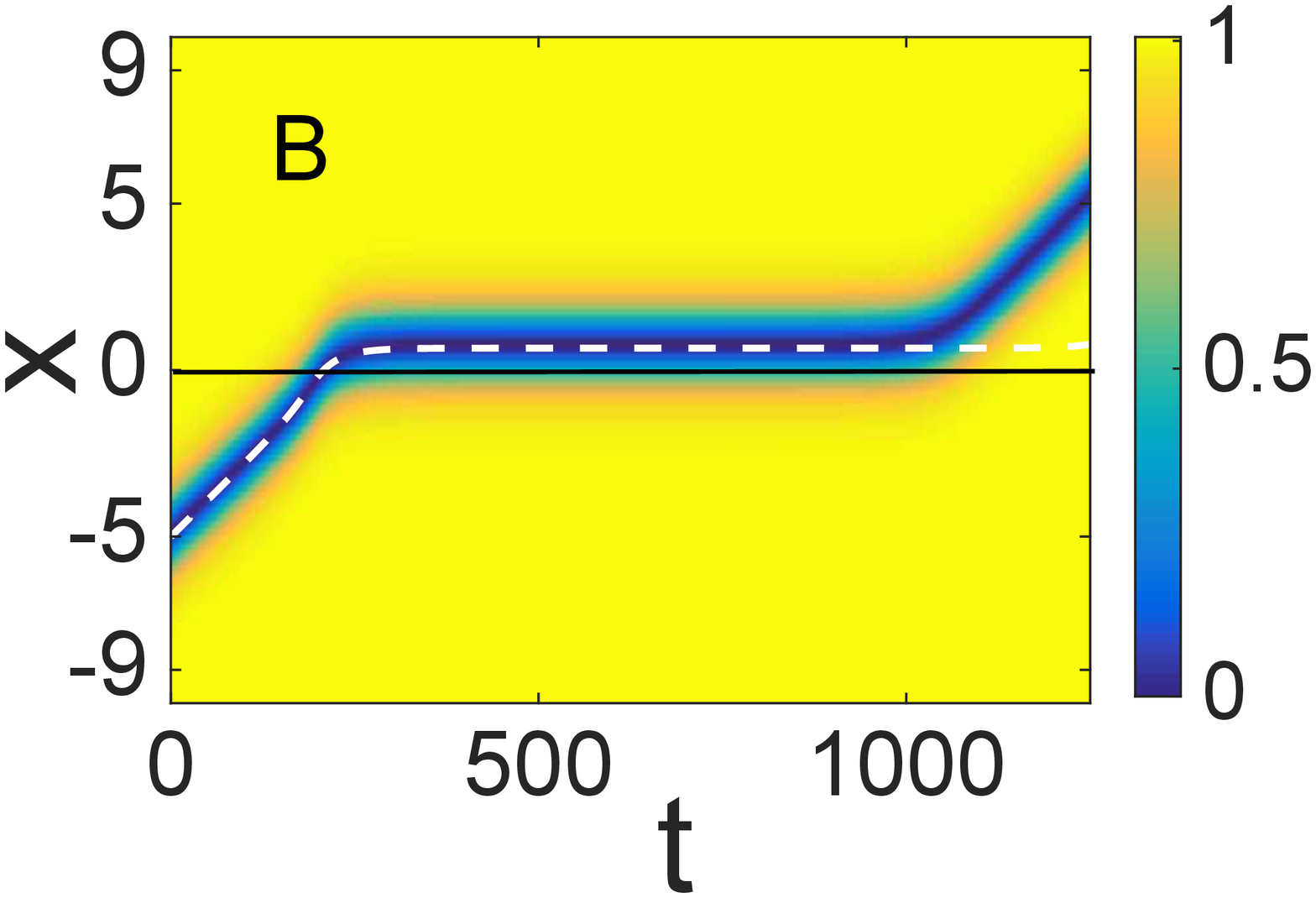} \\
\includegraphics[scale=0.195]{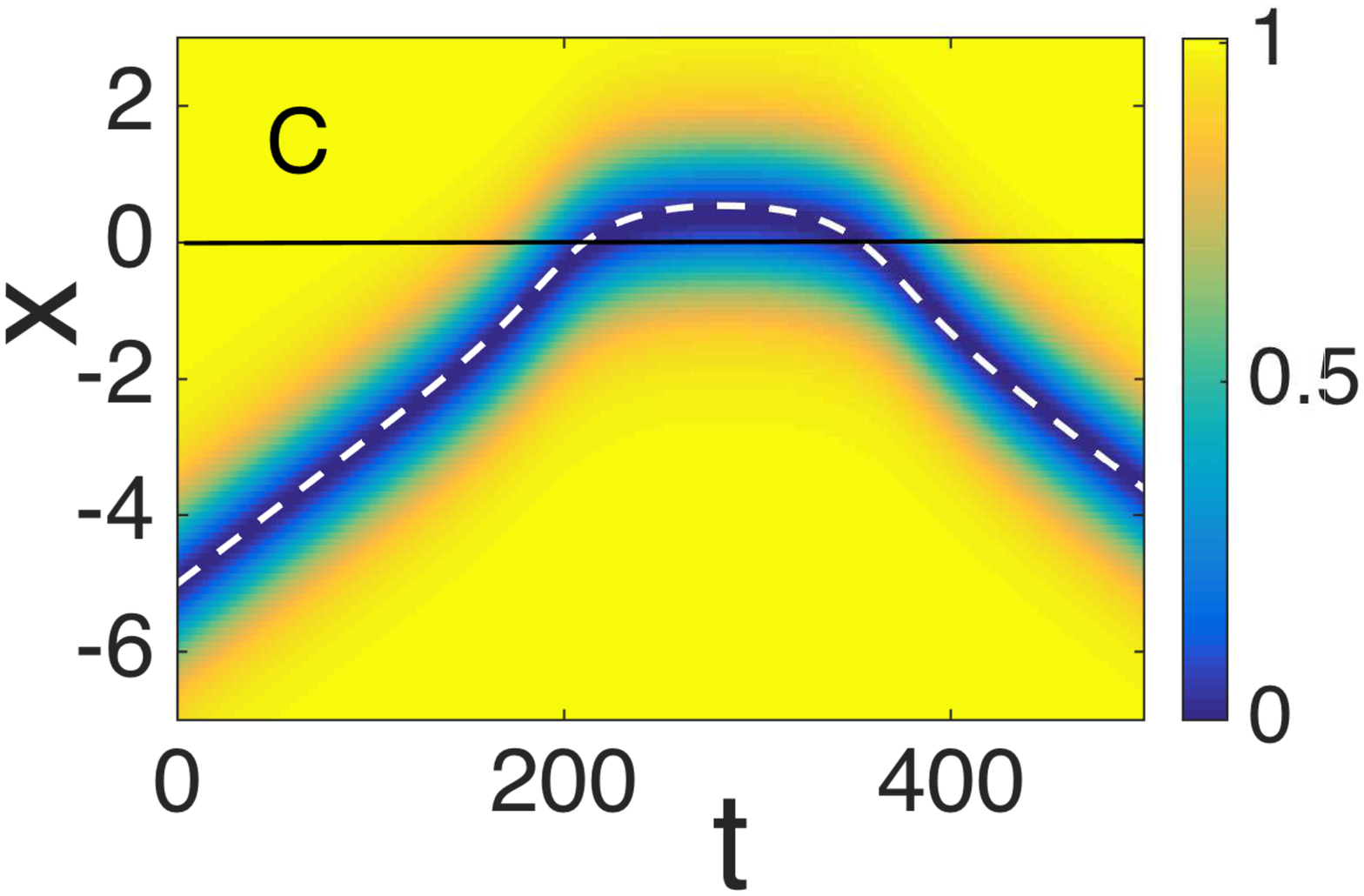}
\includegraphics[scale=0.195]{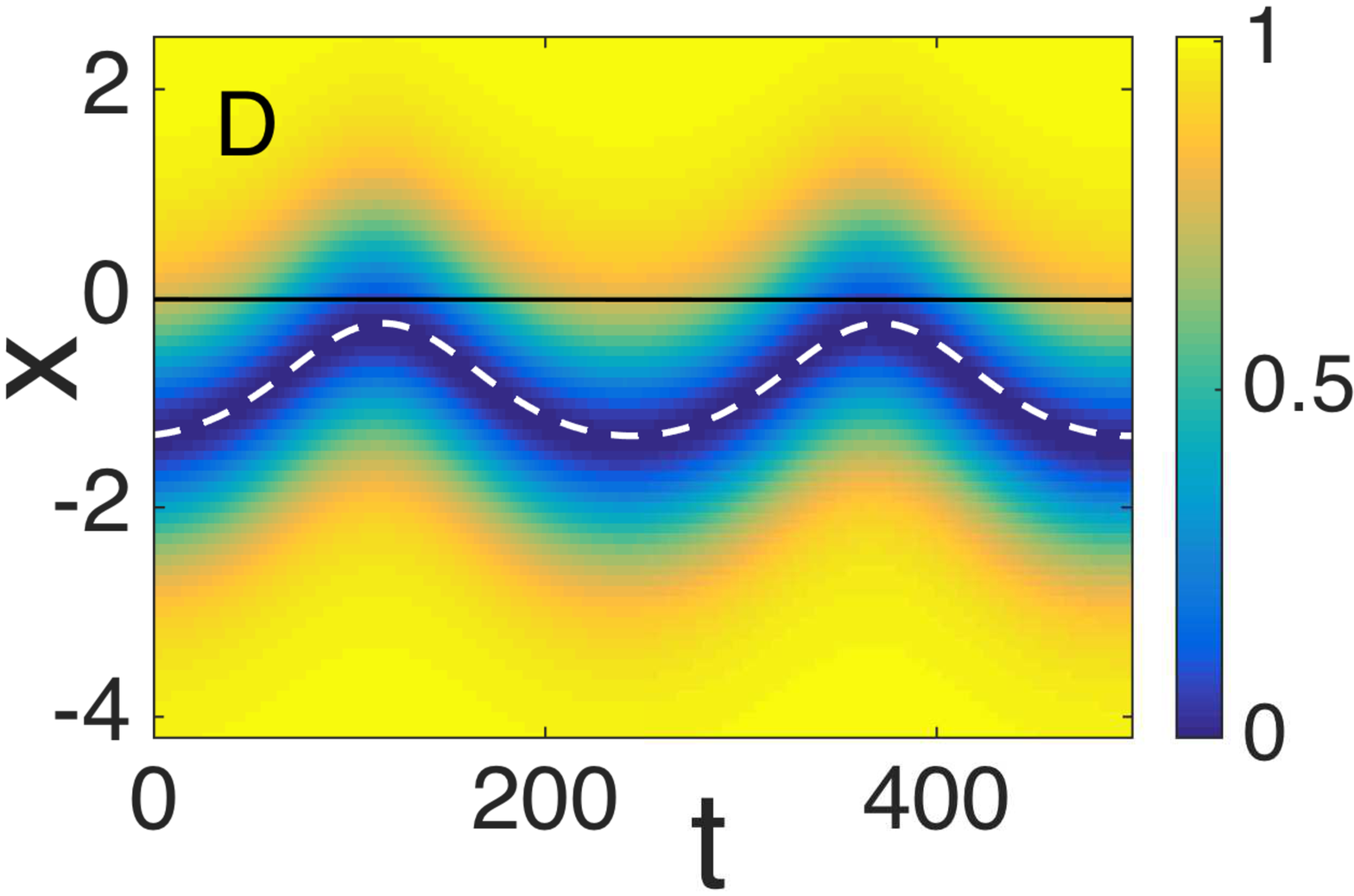}
\caption{(Color online) Similar to Fig.~\ref{ref_tr}, but for a potential
and a nonlinearity step, $A=0.01$ and $B=-0.015$, corresponding to $V_{\rm{L}}=0$,
$V_{\rm{R}}=0.01$, $\alpha_{\rm{R}}/\alpha_{\rm{L}}=0.985$,
and $\mu_{\rm{L}}=1$. Top and bottom panels show the effective potential $W(x_0)$ and
the associated phase plane, respectively; the potential now features
an elliptic and a hyperbolic fixed point at $x_0\approx \pm 0.65$ (cf. vertical dashed lines).
In the phase plane, initial conditions --marked with red squares-- at points
${\rm A}$ ($x_0=-5$, $\phi=0.034$), ${\rm B}$ ($x_0=-5$, $\phi=0.022$), ${\rm C}$
($x_0=-5$, $\phi=0.021$) and ${\rm D}$ ($x_0=-1.3$, $\phi=0.002$)
lead to soliton transmission, quasi-trapping, reflection, and oscillations around
the elliptic fixed point, respectively; asterisks, crosses and stars depict PDE results.
The four bottom respective contour plots show the evolution of the soliton density;
again, thick blue lines and white dashed lines depict PDE and ODE results, respectively.
}
\label{pp1}
\end{figure}

Next, we study the case where both a potential and a nonlinearity step are present (i.e., $A,B \ne 0$),
and there exist fixed points of the effective potential. One such case that we consider in more detail
below is the one corresponding to $A=0.01$ and $B=-0.015$ (respective parameter values are
$V_{\rm{L}}=0$, $V_{\rm{R}}=0.01$, $\alpha_{\rm{R}}/\alpha_{\rm{L}}=0.985$, and $\mu_{\rm{L}}=1$).
Note that for this choice the effective potential asymptotically vanishes,
as shown in the top panel of Fig.~\ref{pp1}; nevertheless, results qualitatively similar to the ones
that we present below can also be obtained for nonvanishing asymptotics of $W(x_0)$.

The effective potential now features an elliptic and a hyperbolic fixed point, located at
$x_0\approx \mp 0.65$ respectively. In this case too, one can identify an
energy threshold $\Delta W$, now defined as $\Delta W = W(x_{0+})-W(-\infty)=W(x_{0+})$,
needed to be overcome by the soliton
kinetic energy in order for the soliton to be transmitted (otherwise, i.e., for $K<\Delta W$, the soliton
is reflected). Using the above parameter values, we find that $\Delta W=2.4\times10^{-4}$ and, hence,
according to Eq.~(\ref{kc}), the critical phase angle for transmission/reflection
is $\phi_{\rm c}\approx 0.022$.
In the simulations, we considered a soliton with initial position and phase angle
$x_0 =-5$ and $\phi=0.034>\phi_{\rm c}$, respectively (cf. point A in the phase plane
shown in the second panel of Fig.~\ref{pp1}), and found that, indeed, the soliton is
transmitted through the effective potential barrier of strength $\Delta W$.
The respective trajectory (starting from point A) is shown in the second
panel of Fig.~\ref{pp1}. Stars along this trajectory, as well as contour plot~A in the same figure,
show PDE results obtained from direct numerical integration of Eq.~(\ref{u});
as in the case of Fig.~\ref{ref_tr}, the (white) dashed line corresponds
to the ODE result.

To study the possibility of soliton trapping, we have also used an initial condition
at the stable branch, incoming towards the hyperbolic fixed point, namely $x_0=-5$ and
$\phi=\phi_{\rm c}\approx 0.022$ (point B in the second panel of Fig.~\ref{pp1}).
In this case, the soliton reaches at the location of the hyperbolic fixed point
(cf. incoming branch, marked with pluses) and appears to be trapped
at the saddle; however, this trapping occurs only for
a finite time (for $t\approx 600$). At the PDE level, this can be understood by the the fact
that such a configuration (i.e., a stationary dark soliton located at the hyperbolic fixed point)
is unstable, as per the analysis of Sec.~II.D. Then, the soliton escapes and moves to the region
of $x>0$, following the trajectory marked with pluses for $x>x_{0+}$ (here,
the pluses depict the PDE results).
The corresponding contour plot~B, in the third panel of Fig.~\ref{pp1}, shows the evolution of the dark
soliton density. Note that, in this case, the result obtained by the ODE (cf. white dashed line) is only
accurate up to the escape time, as small perturbations within the
infinite-dimensional system destroy the delicate balance of the unstable
fixed point.

For the same form of the effective potential, we have also used initial conditions that
lead to soliton reflection. In particular, we have again used
$x_0=-5$ and $\phi=0.021<\phi_{\rm c}$, as well as
an initial soliton location closer to the potential and nonlinearity step, namely $x_0=-1.3$,
and $\phi=0.002$. These initial conditions are respectively indicated by the (red) squares
C and D in the second panel of Fig.~\ref{pp1}. Relevant trajectories in the phase plane,
as well as respective PDE results (cf. stars and X marks), can also be found in the same panel,
while contour plots C and D in the bottom panel of Fig.~\ref{pp1} show the evolution
of the soliton densities. It can readily be observed that for the slightly subcritical value of
the phase angle ($\phi=0.021$), the soliton is again quasi-trapped at the hyperbolic fixed point, but for a
significantly smaller time (for $t\approx 150$). On the other hand, when the soliton
is initially located closer to the steps and has a sufficiently small initial velocity,
it performs oscillations, following the periodic orbit shown in the second panel of Fig.~\ref{pp1}.

\begin{figure}[tbp]
\centering
\includegraphics[scale=0.38]{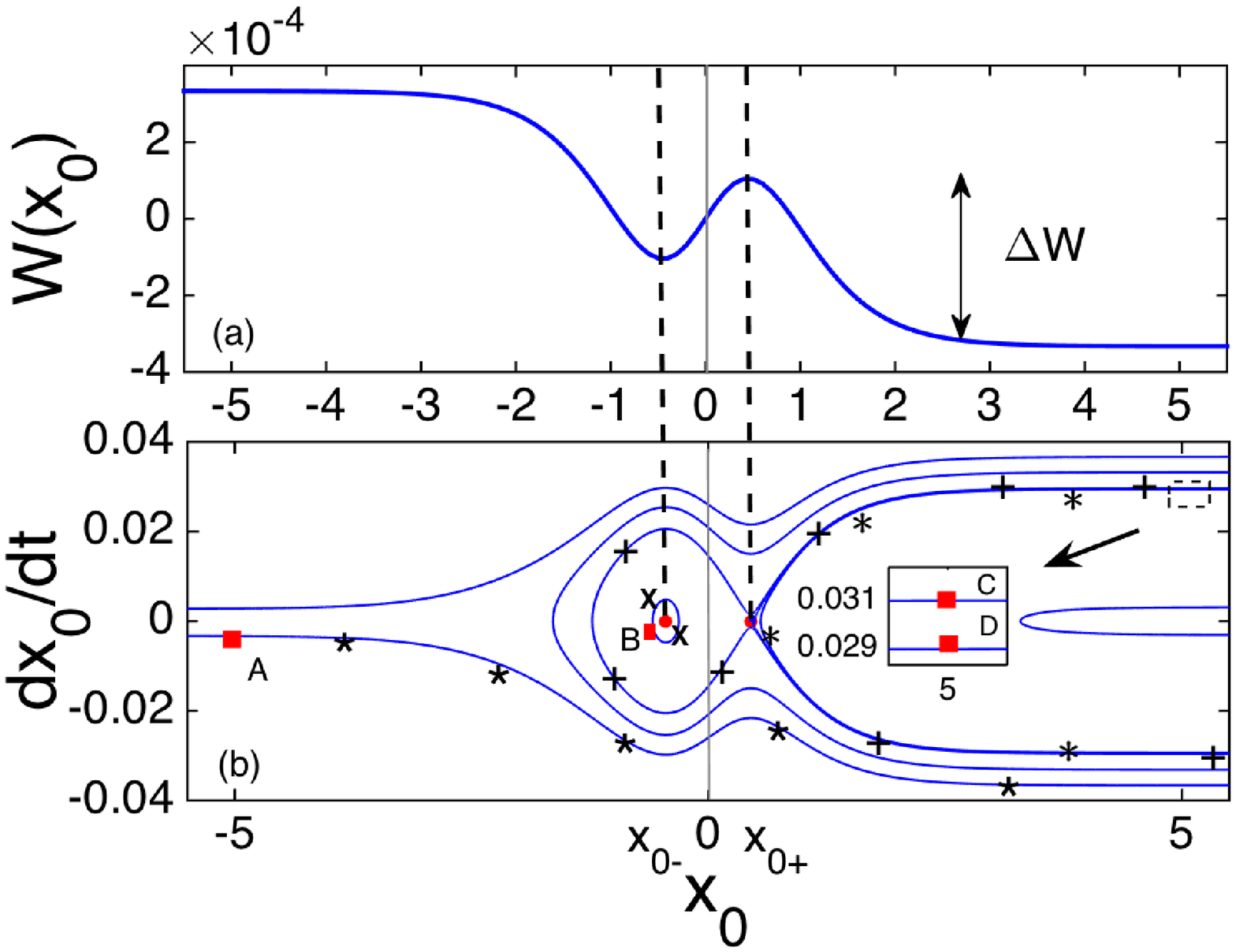}
\includegraphics[scale=0.195]{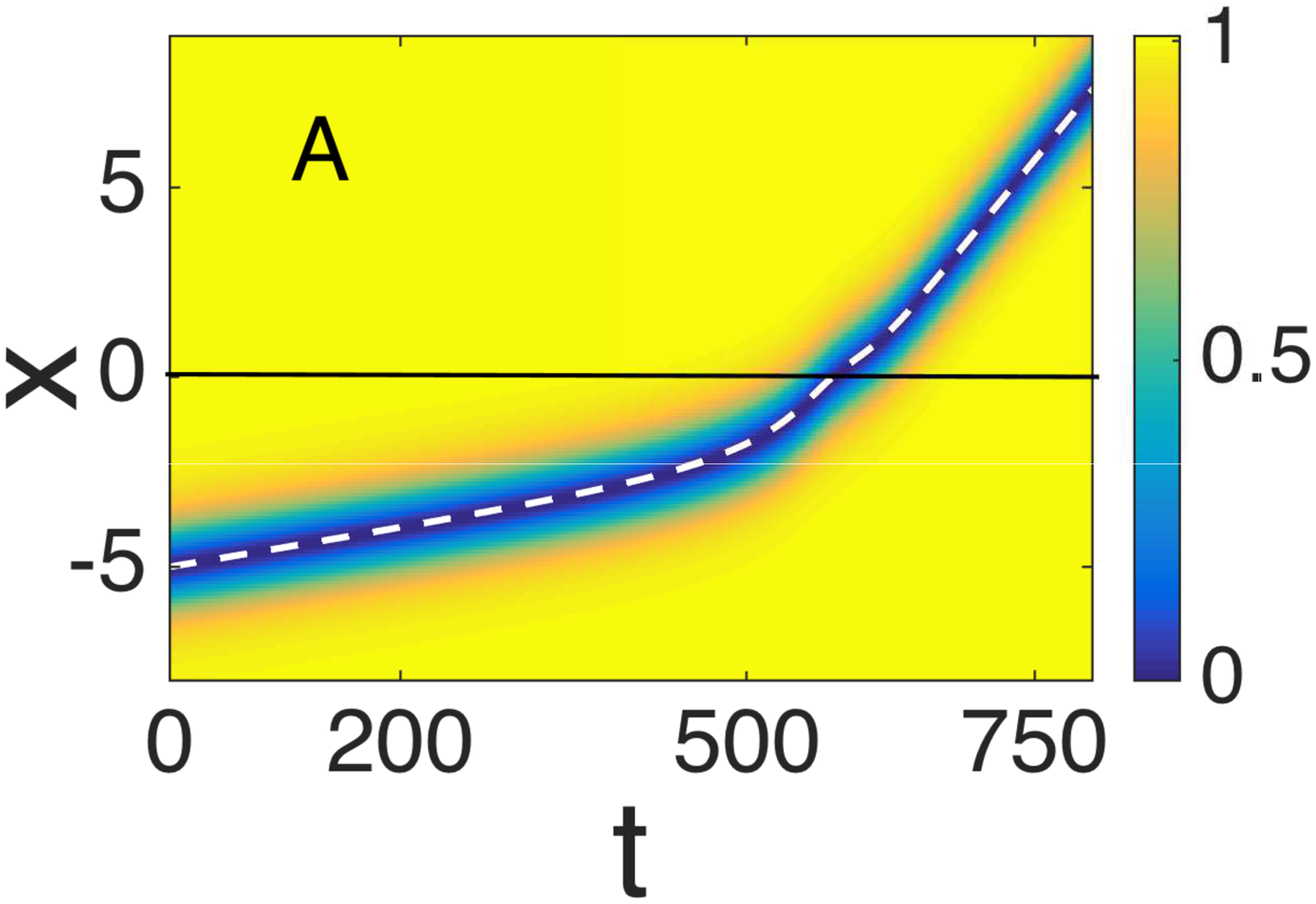}
\includegraphics[scale=0.195]{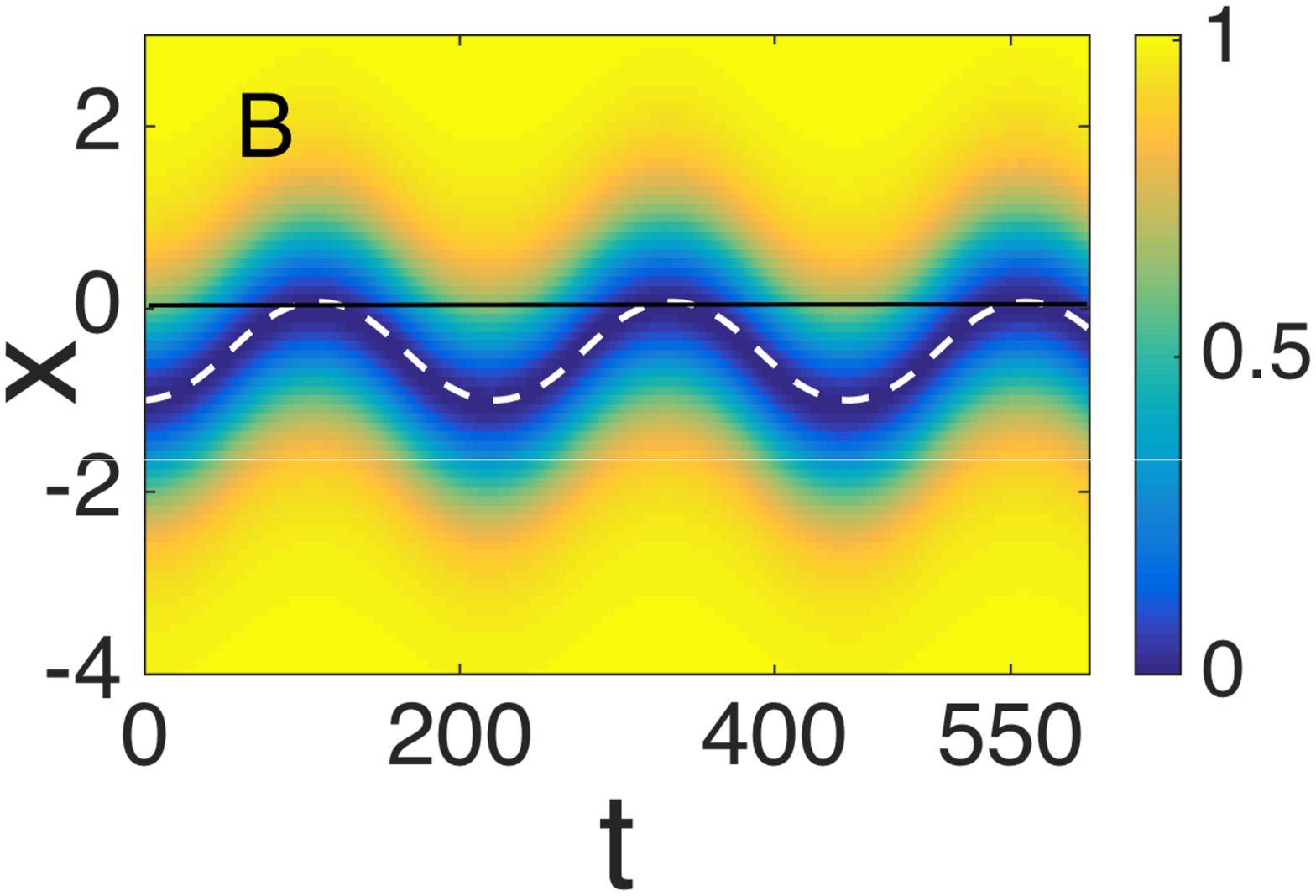}
\includegraphics[scale=0.182]{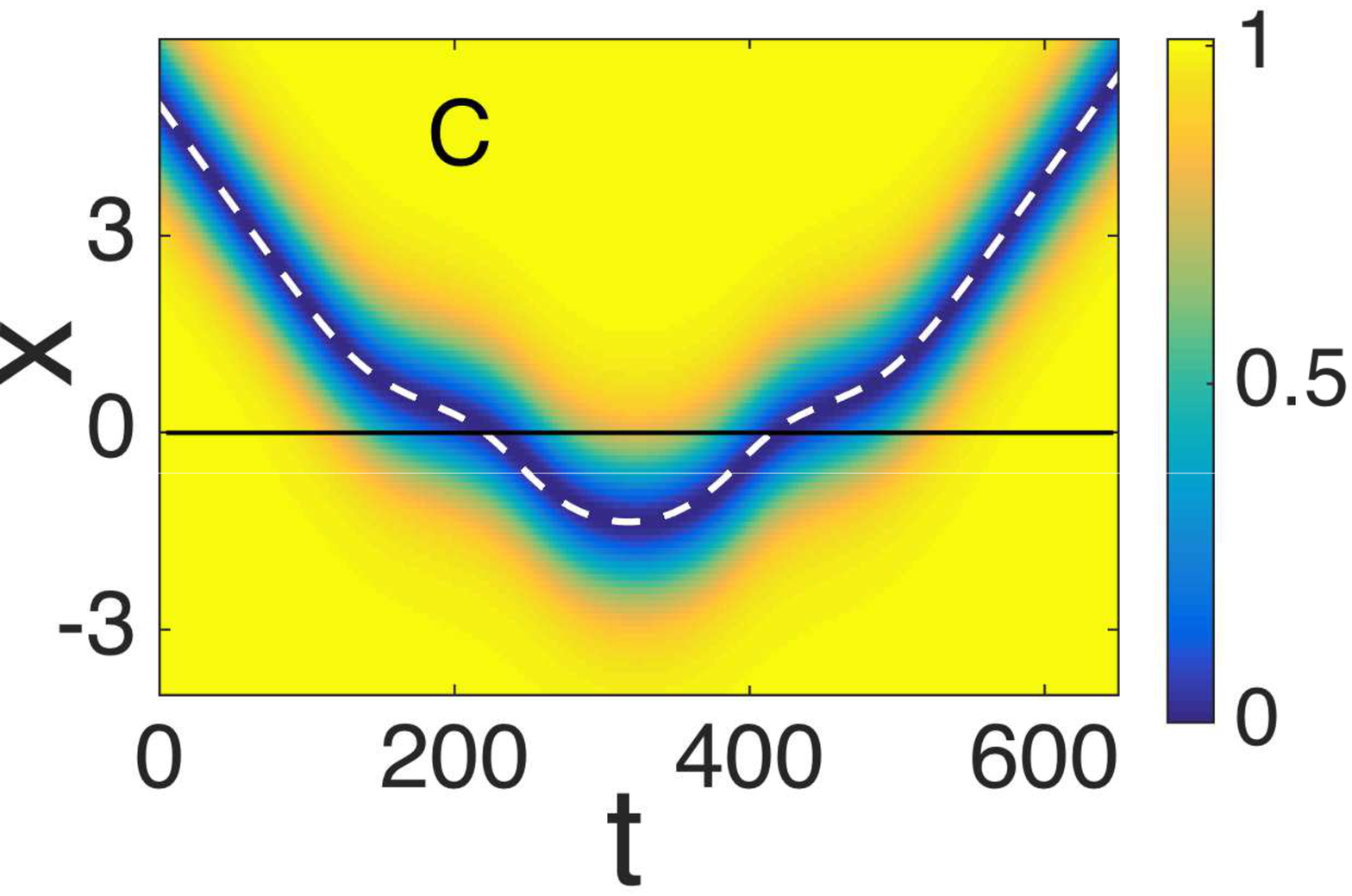}
\includegraphics[scale=0.182]{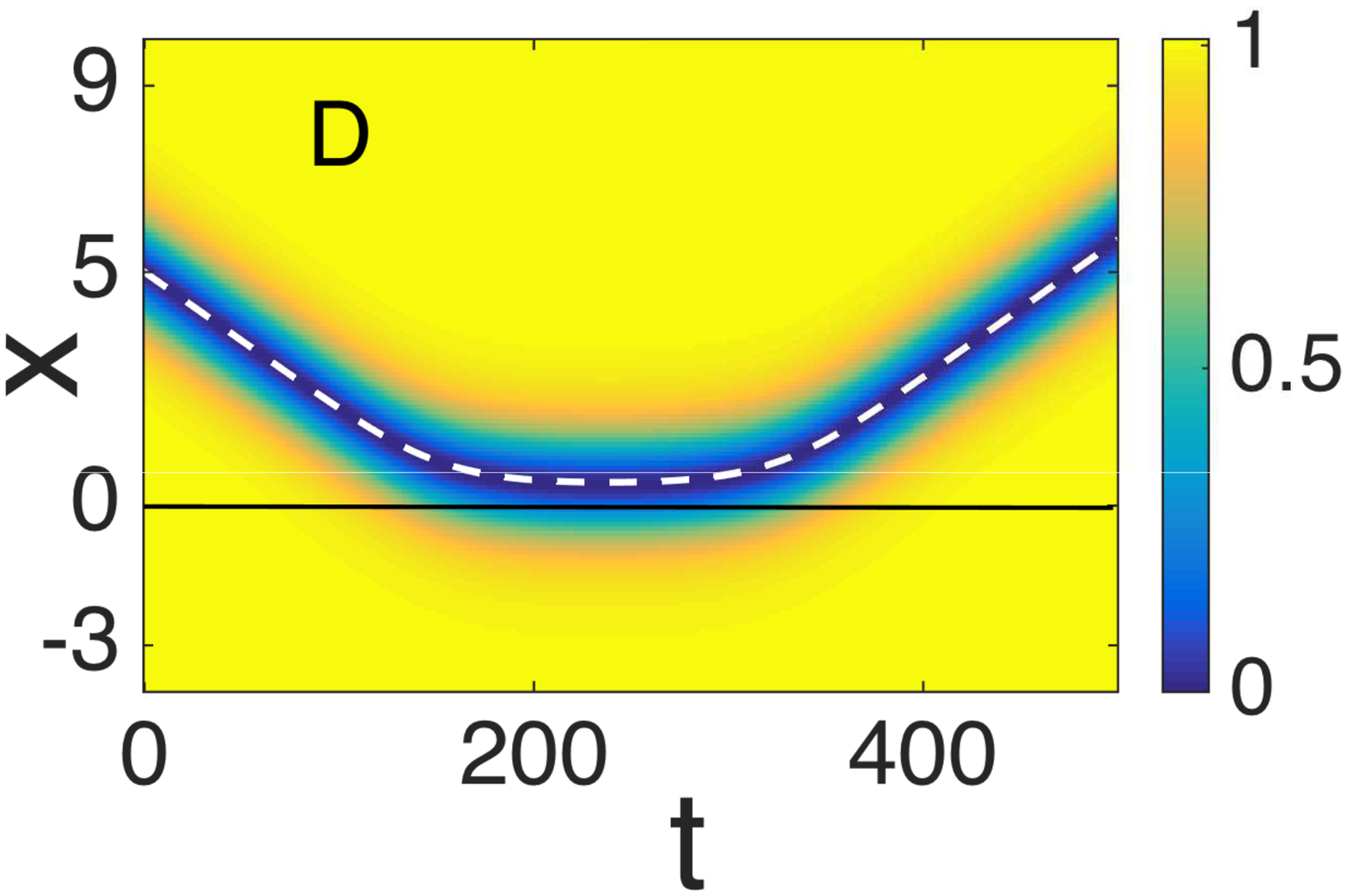}
\caption{(Color online) Similar to Fig.~\ref{pp1}, for a potential and a nonlinearity step, but now
for $A=0.01$ and $B=-0.017$, corresponding to $V_{\rm{L}}=0$, $V_{\rm{R}}=0.01$,
$\alpha_{\rm{R}}/\alpha_{\rm{L}}=0.983$, and $\mu_{\rm{L}}=1$.
The effective potential $W(x_0)$ (top panel), exhibits an elliptic and a hyperbolic fixed point,
at $x_{0\pm}=\pm 0.44$ (vertical dashed lines). In the associated phase plane (second panel) shown are
initial conditions, for a soliton moving to the right, at points A ($x_0=-5$, $\phi=0.005$) and B
($x_0=-1$, $\phi=0.001$), as well as for a soliton moving to the left, at points C
($x_0=5$, $\phi=0.031>\phi_{\rm c}\approx 0.030$) and D ($x_0=5$, $\phi=0.029<\phi_{\rm c}$);
in the relevant trajectories, stars, X marks, pluses and asterisks, respectively, denote PDE results.
Corresponding contour plots for the soliton density are shown in the bottom panels, with the dashed white
lines depicting ODE results.
}
\label{pp2}
\end{figure}

In all the above cases, we find a very good agreement between the analytical predictions
and the numerical results. Similar agreement was also found for other forms of the
effective potential, as shown, e.g., in the example of Fig.~\ref{pp2}
(see also inset~III of Fig.~\ref{exis}). For this form of $W(x_0)$, parameters $A$ and $B$ are
$A=0.01$ and $B=-0.017$ (for $V_{\rm{L}}=0$, $V_{\rm{R}}=0.01$,
$\alpha_{\rm{R}}/\alpha_{\rm{L}}=0.983$, and $\mu_{\rm{L}}=1$), while there
exist again an elliptic and a hyperbolic fixed point,
at $x_{0\pm}=\pm 0.44$ respectively. In such a situation,
if a soliton moves from the left towards the potential and nonlinearity steps,
and is placed sufficiently far from (close to) them -- cf. initial condition at point A
(point B) -- then it will be transmitted (perform oscillations around $x_{0-}$).
On the other hand, if a soliton is initially
placed at some $x_0>x_{0+}$ and moves to the left towards the potential and nonlinearity steps,
it faces an effective barrier $\Delta W$ (cf. top panel of Fig.~\ref{pp2}), now defined as
$\Delta W =W(x_{0+})-W(+\infty)$. In this case too, choosing an an initial condition corresponding
to the stable branch, incoming towards $x_{0+}$, i.e., for the critical phase angle $\phi_{\rm c}\approx 0.03$,
it is possible and achieve quasi-trapping of the soliton for a finite time,
of the order of $t\approx 600$. As such a situation was already discussed above
(cf. panel~B of Fig.~\ref{pp1}), here we present results pertaining to the slightly
supecritical and subcritical cases, namely $\phi=0.031>\phi_{\rm c}$ and $\phi=0.029<\phi_{\rm c}$;
cf. (red) squares C and D in the second panel, and corresponding contour plots in the bottom panel
of Fig.~\ref{pp2}. It is readily observed that, in the former case, the soliton is initially transmitted
through the interface; however, it then follows a trajectory
surrounding the homoclinic orbit (see the orbit marked with plus
symbols, which depicts the PDE results,
in the second panel of Fig.~\ref{pp2}), and is eventually reflected. In the case
$\phi=0.029<\phi_{\rm c}$, the soliton reaches $x_{0+}$, stays there for
a time $t \approx 180$, and eventually is reflected back following the
trajectory marked with asterisks (see second panel of Fig.~\ref{pp2}).
In all cases pertaining to this form of $W(x_0)$, the agreement between
the analytical predictions and the numerical results is very good as well.


\subsection{Rectangular barriers}

Our analytical approximation can straightforwardly be extended to the case of multiple
potential and nonlinearity steps. Here, we will present results for such
a case, where two steps, located at $x=-L$ and $x=L$, are combined so as to form
rectangular barriers, in both the linear potential and the nonlinearity of the system.
In particular, we consider the following profiles for the potential and the scattering length:
\begin{eqnarray}
V(x)&=&V_b(x)+
\begin{cases}
V_{\rm{R}}, & |x|>L \\
V_{\rm{L}}, & |x|<L \\
\end{cases}, \\
\alpha(x)&=&
\begin{cases}
\alpha_{\rm{R}}, &  |x|>L \\
\alpha_{\rm{L}}, & |x|<L
\end{cases}.
\label{setting2}
\end{eqnarray}
In such a situation, the effective potential can be found following the lines of the
analysis presented in Sec.~II.B: taking into regard that the perturbation $P(\upsilon)$
in Eq.~(\ref{upsilon}) has now the form:
\begin{eqnarray}
P(\upsilon)&=&\left(A+B |\upsilon|^2\right)\upsilon \left[\mathcal{H}(x+L)-\mathcal{H}(x-L)\right],
\label{P2}
\end{eqnarray}
it is straightforward to find that the relevant effective potential is given by:
\begin{eqnarray}
W(x_0) &=& \frac{1}{8}\big( 2A + B \big)
[\tanh(L-x_0)+\tanh(L+x_0)] \nonumber \\
&+&\frac{1}{24} B [\tanh^3(L-x_0) +\tanh^3(L+x_0)].
\label{W2}
\end{eqnarray}
\begin{figure}[tbp]
\centering
\includegraphics[scale=0.38]{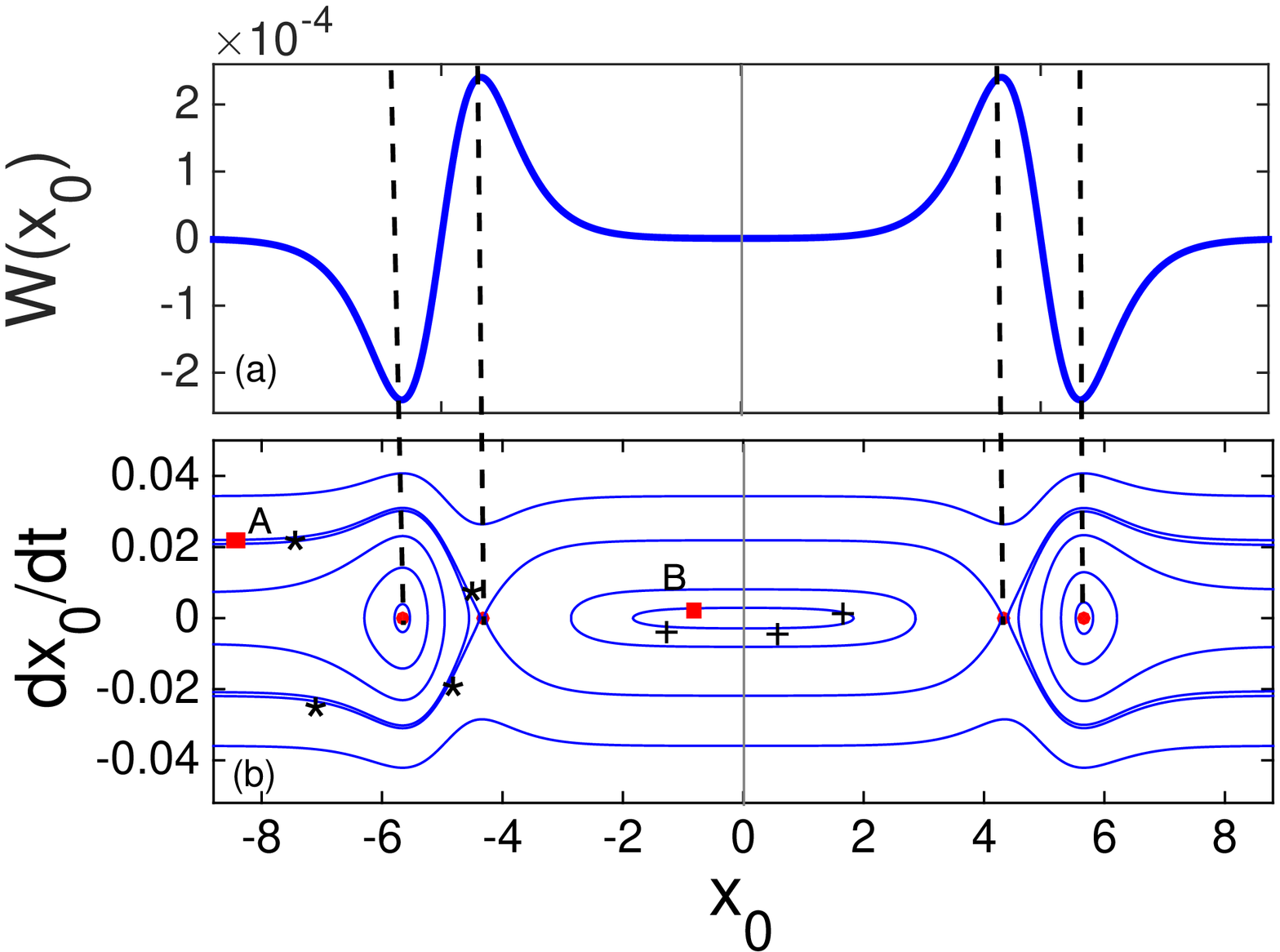}
\includegraphics[scale=0.195]{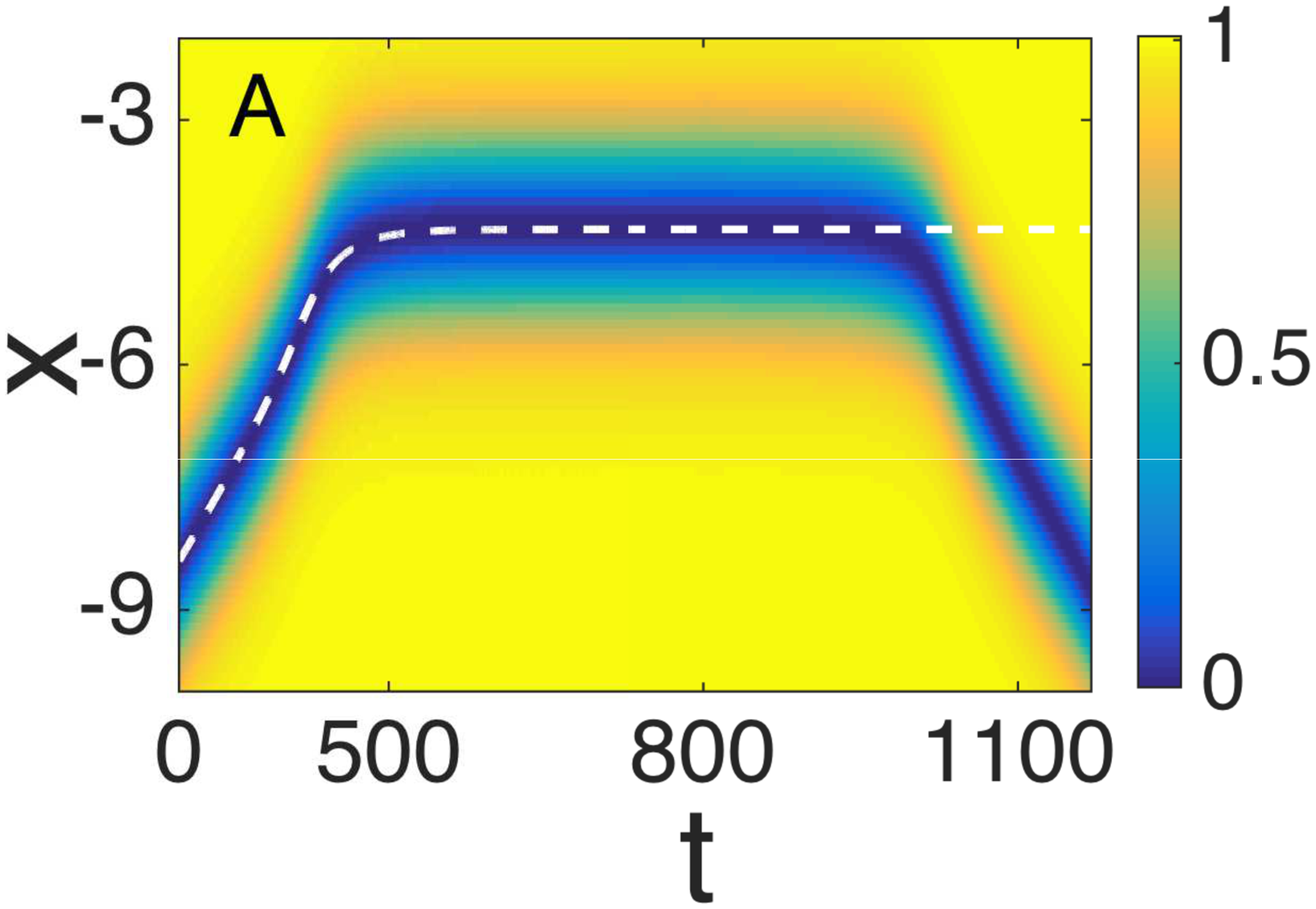}
\includegraphics[scale=0.195]{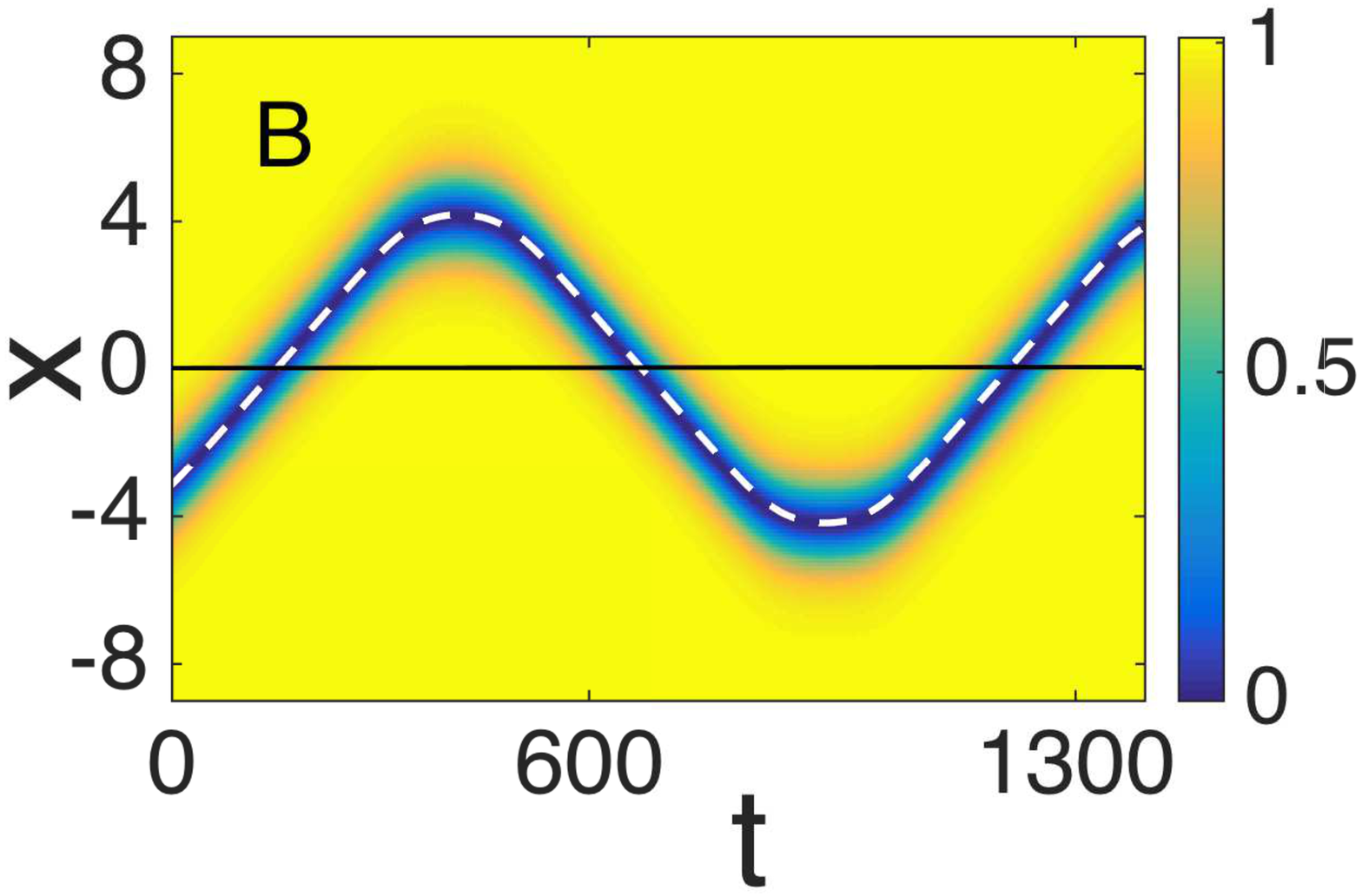}
\caption{(Color online) The case of two potential and nonlinearity steps forming respective
rectangular barriers, for $L=5$, $A=0.01$ and $B=-0.015$, corresponding to
$V_{\rm{L}}=0$, $V_{\rm{R}}=0.01$, $\alpha_{\rm{R}}/\alpha_{\rm{L}}=0.985$, $\mu_{\rm{L}}=1$.
Top panel (a): the effective potential $W(x_0)$ [cf. Eq.~(\ref{W2})], featuring
elliptic fixed points at the origin and at $\pm 5.66$, and
a pair of hyperbolic fixed points at $\pm 4.34$.
Middle panel (b): the associated phase plane; (red) squares
${\rm A}$ and ${\rm B}$ depict different initial conditions, corresponding to quasi-trapping
or oscillations, while stars and crosses depict respective PDE results.
Bottom panels: contour plots showing the evolution of the dark soliton density
for the initial conditions depicted in
the middle panel, i.e., $x_0=-8.6$ and $\phi=2.2\times 10^{-2}$ (left),
or $x_0=-3$ and $\phi=3\times 10^{-3}$ (right); here, as before,
dashed (white) lines depict ODE results.
}
\label{step}
\end{figure}

Typically, i.e., for sufficiently large arbitrary values of $L$, the effective potential
is as shown in the top panel of Fig.~\ref{step}; in this example, we used $L=5$, while
$A=0.01$ and $B=-0.015$. It is readily observed that, in this case, associated with
such a potential and a nonlinearity barrier, is an effective potential of the form of
a superposition of the ones shown in Fig.~\ref{pp1}, which are now located at $\pm 5$.
The associated phase plane is shown in the middle panel of Fig.~\ref{step}; shown also are
initial conditions corresponding to soliton quasi-trapping,
or oscillations around the elliptic fixed point at the origin
-- cf. red square points A and B, respectively.
The corresponding soliton trajectories are depicted both in the phase plane
in the middle panel of Fig.~\ref{step} and in the space-time contour plots
showing the evolution of the soliton density in the bottom panels of the same figure.
Note that stars and plus symbols in the middle panel correspond
to PDE results, obtained in the framework of Eq.~(\ref{u}),
while the (white) dashed lines in the bottom panels depict ODE results,
obtained by Eq.~(\ref{eqmotion}) for the potential in Eq.~(\ref{W2}. Obviously,
once again, agreement between theoretical predictions and numerical results is very good.

An interesting situation occurs as $L$ decreases. To better illustrate what happens
in this case, and also to make connections with earlier work \cite{gt}, we consider
the simpler case of $B=0$ (i.e., the nonlinearity step is absent). Then, assuming that
$A=b/(2L)$ (with $b$ being an arbitrary small parameter), and in the limit
of $L\rightarrow 0$, the potential step takes the form of a delta-like impurity of strength $b$.
In this case, the effective potential of Eq.~(\ref{W2}) is reduced to the form
$W(x_0)=(b/4){\rm sech}^2(x_0)$. This result recovers the one reported in Ref.~\cite{gt}
(see also Refs.~\cite{vvk1,nn}), where
the interaction of dark solitons with localized impurities was studied; cf.
Eq.~(16) of that work, but in the absence of the trapping potential $U_{\rm tr}$.

\begin{figure}[tbp]
\centering
\includegraphics[scale=0.35]{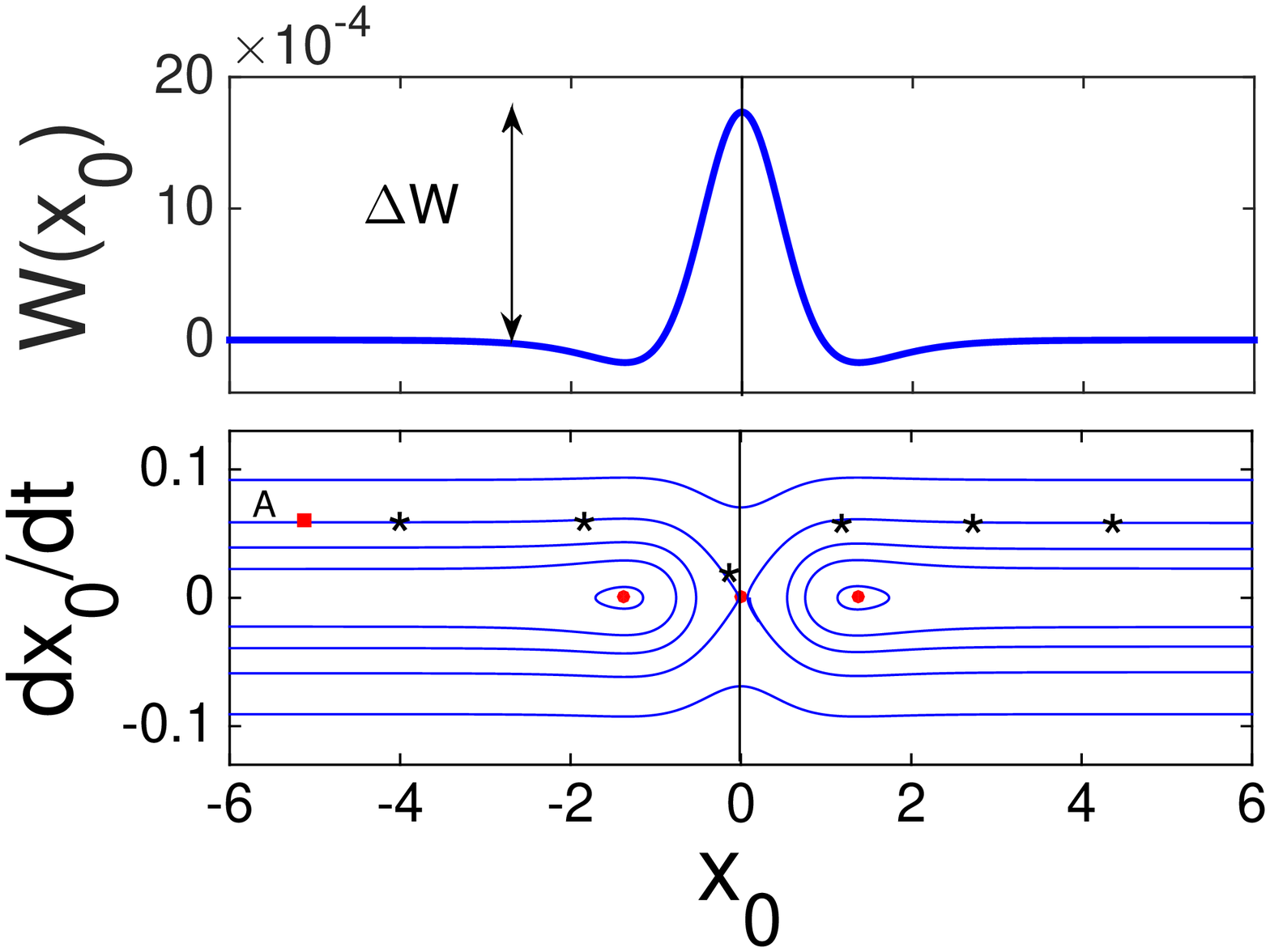}
\includegraphics[scale=0.23]{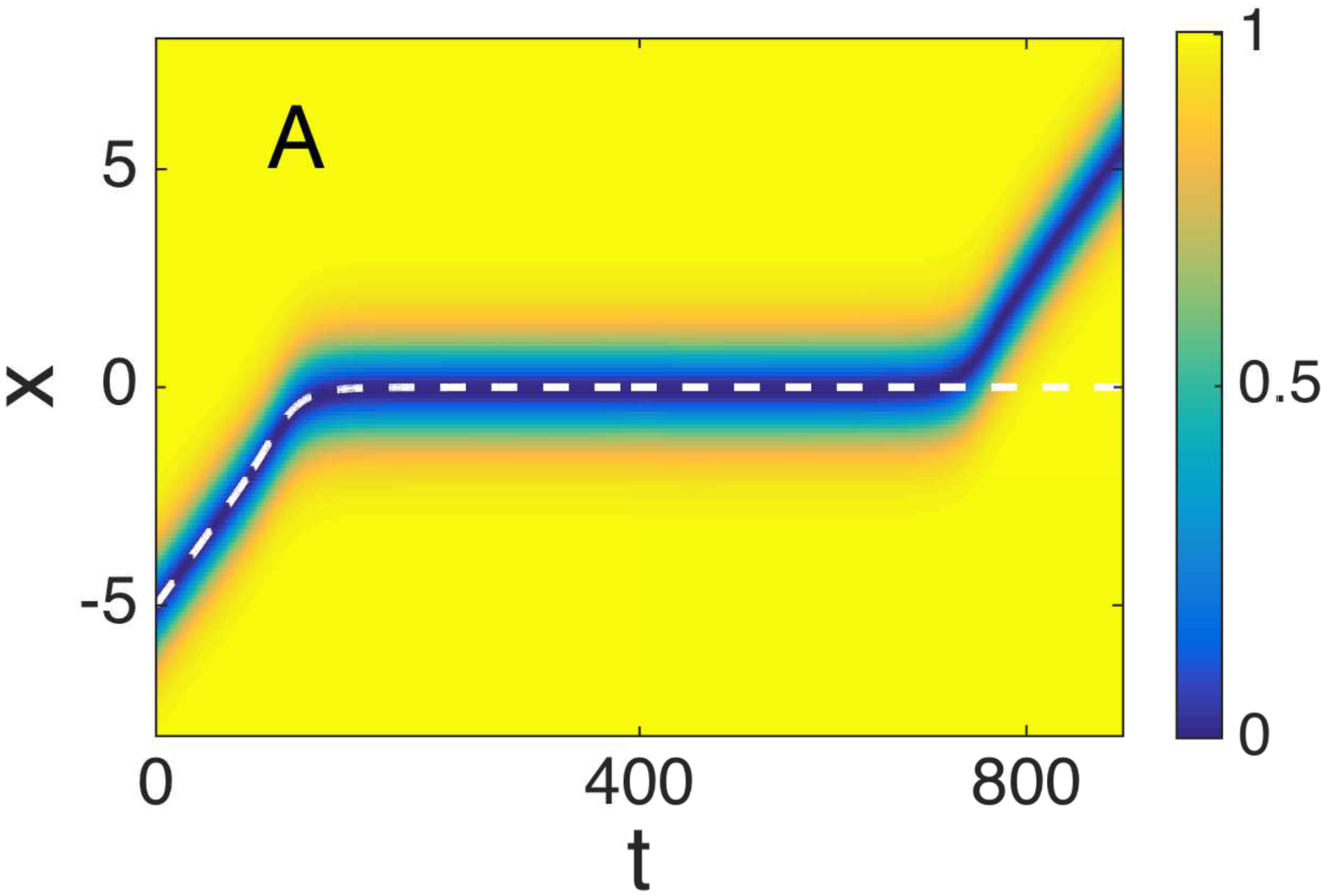}
\caption{(Color online) The case of two potential and nonlinearity steps forming respective
rectangular barriers, for $L=0.1$, $A=0.1$ and $B=-0.13$, corresponding to $a_{\rm R}/a_{\rm L}=0.87$,
$V_{\rm R}=0.1$ and $\mu_{\rm L}=1$. Top panel: the effective potential $W(x_0)$,
featuring a hyperbolic fixed point at the origin and a pair of elliptic fixed points at $\pm 1.38$.
Middle panel: the associated phase plane; (red) square
${\rm A}$ depicts an initial condition corresponding to quasi-trapping of the soliton,
while stars depict respective PDE results.
Bottom panel: contour plot showing the evolution of the dark soliton density
for the initial condition depicted in
the middle panel, i.e., $x_0=-5$ and $\phi=5.8\times 10^{-2}$; as before,
dashed (white) lines depict ODE results.
}
\label{delta2}
\end{figure}

In the same limiting case of small $L$, and for $B\ne 0$, the effective potential
has typically the form shown in the top panel of Fig.~\ref{delta2}; here, we use $L=0.1$,
while $A=0.1$ and $B=-0.13$, corresponding to $a_{\rm R}/a_{\rm L}=0.87$, $V_{\rm R}=0.1$
and $\mu_{\rm L}=1$.
Comparing this form of $W(x_0)$ with the one shown in Fig.~\ref{step}, it becomes clear
that as $L\rightarrow 0$, the individual parts of the effective potential of Fig.~\ref{step}
pertaining to the two potential/nonlinearity steps move towards the origin. There, they merge
at the location of the ``central'' elliptic fixed point, which becomes unstable through
a pitchfork bifurcation. As a result of this process, an unstable (hyperbolic)
fixed point emerges at the origin, while the ``outer'' pair of the elliptic fixed points
(cf. Fig.~\ref{step}) also drift towards the origin -- in this case, they are located at $\pm 1.38$.

In the middle panel of Fig.~\ref{delta2}, shown also is the phase plane associated to
the effective potential of the top panel. As in the cases studied in the previous sections,
we may investigate possible quasi-trapping of the soliton, using an initial condition at
the stable branch, incoming towards the hyperbolic fixed point at the origin. Indeed, choosing
$x_0=-5$ and $\phi=\phi_c=5.8\times 10^{-2}$ (notice that here, the corresponding effective barrier
$\Delta W = 1.7\times 10^{-3}$ -- cf. top panel of Fig.~\ref{delta2}), we find the
following: the soliton arrives at the origin, stays there for a time $t \approx 600$,
and then it is transmitted through the region $x>0$. In fact, the corresponding trajectory
found at the PDE level is depicted by stars in the middle panel of Fig.~\ref{delta2},
while the relevant contour plot showing the evolution of the soliton density is shown in the
bottom panel of the same figure. Notice, again, the fairly good agreement between
numerical and analytical results.

We note that for the same parameter values, but for $B=0$, elliptic fixed points
do not exist, and the effective potential has simply the form of a sech$^2$ barrier,
as mentioned above (see also work of Ref.~\cite{gt}). In this case, starting
from the same initial position, $x_0=-5$, and for $\phi=0.1$ (corresponding to
$\phi_c =\sqrt{2\Delta W} \approx 0.1$), we find that the trapping time is $t\approx 320$,
i.e., almost half of the one that was when the nonlinearity steps are present
(results not shown here). This observation, along with the results presented in the
previous sections, indicate that nonlinearity steps/barriers are necessary
either to facilitate or enhance soliton trapping in such inhomogeneous settings.

\section{Discussion and conclusions}

We have studied matter-wave dark solitons near linear
potential and nonlinearity steps,
superimposed on a box-like potential that was assumed to confine the atomic
Bose-Einstein condensate. The formulation of the problem finds a direct application
in the context of nonlinear optics: the pertinent model can be used to describe the
evolution of beams, carrying dark solitons, near interfaces separating optical media
with different linear refractive indices and different defocusing Kerr nonlinearities.

Assuming that the potential/nonlinearity steps were small, we employed perturbation
theory for dark solitons to show that, in the adiabatic approximation, solitons behave
as equivalent particles moving in the presence of an effective potential. The latter was found
to exhibit various forms, ranging from simple tanh-shaped steps -- for a spatially homogeneous
scattering length (or same Kerr nonlinearity, in the context of optics) -- to more complex forms, featuring hyperbolic and elliptic fixed points -- in the presence of steps in the scattering length
(different Kerr nonlinearities).

In the latter case, we found that stationary soliton states do exist at the fixed points of
the effective potential. Using a Bogoliubov-de Gennes (BdG) analysis, we showed that these states
are unstable: dark solitons at the hyperbolic fixed points have a pair of unstable real eigenvalues,
while those at the elliptic fixed points have a complex eigenfrequency quartet, dictating
a purely exponential or an oscillatory instability, respectively. We also used an analytical
approximation to determine the real and imaginary parts of the relevant eigenfrequencies as
functions of the nonlinearity step strength. The analytical predictions were found to be
in good agreement with corresponding numerical findings obtained in the framework of the
BdG analysis.

We then studied systematically soliton dynamics, for a variety of parameter values
corresponding to all possible forms of the effective potential.
Adopting the aforementioned equivalent particle picture, we found analytically
necessary conditions for soliton reflection at, or transmission through
the potential and nonlinearity steps: these correspond to initial soliton
velocities smaller or greater to the energy of the effective steps/barriers
predicted by the perturbation theory and the equivalent particle picture.

We also investigated
the possibility of soliton (quasi-) trapping, for initial conditions corresponding to the incoming, stable
manifolds of the hyperbolic fixed points (which exist only for inhomogeneous nonlinearities).
In the context of optics, such a trapping can be regarded as the formation of surface dark solitons
at the interface between dielectrics of different refractive indices.
We found that trapping is possible, but only for a finite time. This effect can be understood by
the fact that stationary solitons at the hyperbolic fixed points
are unstable, as was
corroborated by the BdG analysis.
Thus small perturbations (at the PDE level) eventually cause the
departure of the solitary wave from the relevant fixed points.
Nevertheless, it should be pointed out that the time of
soliton quasi-trapping was found to be of the order of $600\sqrt{2}\xi/c_S$ in physical units;
thus, typically, for a healing length $\xi$ of the order of a micron and a speed of sound
$c_s$ of the order of a millimeter-per-second, trapping time may be of the order of
$\approx 850$~ms. This indicates that such a soliton quasi-trapping effect may be observed
in real experiments. Note that in all scenarios (reflection, transmission, quasi-trapping)
our analytical predictions were found to be in very good agreement with direct numerical
simulations in the framework of the original Gross-Pitaevskii model.

We have also extended our considerations to study cases involving two potential and nonlinearity
steps, that are combined so as to form corresponding rectangular barriers. Reflection,
transmission and quasi-trapping of solitons in such cases were studied too, again with a
very good agreement between analytical and numerical results. In this setting,
special attention was paid to the limiting case of infinitesimally
small distance
between the adjacent potential/nonlinearity steps that form the barriers. In this case,
we found that, due to a pitchfork bifurcation, the stability of the fixed point
of the effective potential at the barrier center changes: out of two
hyperbolic and one elliptic
fixed point, a hyperbolic fixed point emerges, and the potential rectangular barrier
is reduced to a delta-like impurity. The latter is described by a sech$^2$ effective potential,
in accordance with the analysis of earlier works \cite{vvk1,gt,nn}.

Our methodology and results pose a number of interesting questions for future studies.
First, it would be interesting to investigate how our perturbative results change as the
potential/nonlinearity steps or barriers become larger, and/or attain more realistic shapes
(including steps bearing finite widths, as well as Gaussian barriers -- cf., e.g., recent work of Ref.~\cite{sch}).
In the same context, a systematic numerical -- and, possibly, also analytical -- study of the
radiation of solitons during reflection or transmission
(along the lines, e.g., of Ref.~\cite{prouk}) should also provide a more complete
picture in this problem.
Furthermore, a systematic study of settings
involving multiple such steps/barriers, and an investigation of the possibility of
soliton trapping therein, would be particularly relevant. In such settings, investigation
of the dynamics of moving steps/barriers could find direct applications to fundamental
studies relevant, e.g., to superfluidity (see, for instance, Ref.~\cite{engels}),
transport of BECs \cite{hulet}, and even Hawking radiation in analog black hole lasers
implemented with BECs \cite{stein}.
Finally, extension of our analysis to higher-dimensional settings,
would also be particularly challenging:
first, in order to investigate transverse excitation effects that are not captured
within the quasi-1D setting, and second to study similar problems with vortices and
other vortex structures. See, e.g., Ref.~\cite{siam}
for a summary of relevant studies in higher-dimensional settings, 
and Ref.~\cite{bpanew} for a recent
example of manipulation/control of vortex patterns and their formation
via Gaussian barriers, motivated by experimentally accessible laser beams.

\section*{Acknowledgments}

The work of F.T. and D.J.F. was partially supported by the Special Account
for Research Grants of the University of Athens. 
The work of F.T. and Z.A.A. was partially supported by Qatar University under the scope of the 
Internal Grant QUUG-CAS-DMSP-13/14-7.
F.T. acknowledges hospitality at
Qatar University, where most of this work was carried out.
The work of P.G.K.
at Los Alamos is partially supported by the US Department of Energy.
P.G.K. also gratefully
acknowledges the support of NSF-DMS-1312856, BSF-2010239, as well as from the US-AFOSR under grant FA9550-12-1-
0332, and the ERC under FP7, Marie Curie Actions, People, International Research Staff Exchange Scheme (IRSES-605096).

\end{document}